# Cholesterol modulates acetylcholine receptor diffusion by tuning confinement sojourns and nanocluster stability

Alejo Mosqueira, Pablo A. Camino and Francisco J. Barrantes[*]

Laboratory of Molecular Neurobiology, Biomedical Research institute (BIOMED), UCA–CONICET, Av. Alicia Moreau de Justo 1600, C1107AFF Buenos Aires, Argentina.

*Address correspondence to: Dr F.J. Barrantes, Laboratory of Molecular Neurobiology, BIOMED UCA-CONICET, Av. Alicia Moreau de Justo 1600, C1107AFF Buenos Aires, Argentina.
E-mail: rtfjb1@gmail.com

The abbreviations used are: BTX, α-bungarotoxin; nAChR, nicotinic acetylcholine receptor; CDx, methyl-β-cyclodextrin; CDx-Chol, cholesterol-methyl-β-cyclodextrin complex; MSD, mean-squared displacement; SPT, single-particle tracking.

## Abstract

Translational motion of neurotransmitter receptors is key for determining receptor number at the synapse and hence, synaptic efficacy. We combine live-cell STORM superresolution microscopy of nicotinic acetylcholine receptor (nAChR) with single-particle tracking, mean-squared displacement (MSD), turning angle, ergodicity, and clustering analyses to characterize the lateral motion of individual molecules and their collective behaviour. nAChR diffusion is highly heterogeneous: subdiffusive, Brownian and, less frequently, superdiffusive. At the single-track level, free walks are transiently interrupted by ms-long confinement sojourns occurring in nanodomains of ~36 nm radius. Cholesterol modulates the time and the area spent in confinement. Turning angle analysis reveals anticorrelated steps with time-lag dependence, in good agreement with the permeable fence model. At the ensemble level, nanocluster assembly occurs in second-long bursts separated by periods of cluster disassembly. Thus, millisecond-long confinement sojourns and second-long reversible nanoclustering with similar cholesterol sensitivities affect all trajectories; the proportion of the two regimes determines the resulting macroscopic motional mode and breadth of heterogeneity in the ensemble population.

**Introduction**

Transmission of chemical signals in the synapse is regulated by the crosstalk between neurotransmitter receptors and scaffolding proteins, lipids, and the cytoskeleton. To understand synaptic physiology, it is necessary to define the supramolecular organization, local dynamics and trafficking of the intervening actors. The available number of neurotransmitter receptors and the time they spend in the post-synaptic region and, more precisely, at the actives sites facing the neurotransmitter releasing areas, directly affects synaptic activity. Comprehension of the various phenomena that determine this spatio-temporal homeostatic balance is key to understanding the function of the synapse in health and disease. One such phenomenon is the motion of receptors in the plasma membrane. How receptors diffuse into and out of the synaptic region, and which mechanisms affect lateral diffusion and maintenance of receptors in the post-synaptic region are fundamental to an understanding of synaptic function.

The superfamily of pentameric ligand-gated ion channels (pLGIC) is a collection of integral membrane proteins expressed in the central and peripheral nervous system, where they perform important functions in cell-surface signalling by transducing the chemical signal contained in the neurotransmitter into rapid ion fluxes at the post-synaptic membrane. One of the best characterized neurotransmitter receptors is the nicotinic acetylcholine receptor (nAChR) protein, the paradigm member of the pLGIC, a hot focus of research for possible intervention in neurological and neuropsychiatric diseases. The membrane lipid environment in which receptors are embedded affects the functional properties and distribution of the protein macromolecules. Cholesterol is an abundant component in the postsynaptic membrane, and there is a plethora of evidence on the variety of modulatory roles exerted by this lipid on the nAChR [1]. Interestingly, the members of the pLGIC superfamily are descendants of an ancestral "proto-channel" which appeared early in phylogenetic evolution, before the prokaryote-eukaryote dichotomy, and although almost all prokaryotes lack cholesterol, the pLGIC exhibit the same sterol-recognition motifs as their eukaryotic counterparts [2, 3]. This remarkable degree of conservation at the molecular scale points to the importance of cholesterol in pLGIC function. In the specific case of the nAChR, we know that cholesterol exhibits preference over other lipid species for the superficial region surrounding the surface of the receptor protein; about 15 cholesterol molecules appear to be located at this region [4]. These cholesterol molecules are in constant exchange with the bulk bilayer cholesterol [5], which is a key regulator of physical membrane properties at large. At the subcellular scale, there is also evidence that nAChRs interact with cholesterol-rich lipid domains ("rafts") *in vitro* and *in vivo* [6-8]. Acute cholesterol depletion reduces the number of receptor domains by accelerating the rate of endocytosis and shifting the internalization to a different endocytic pathway involving the small GTPase Arf6 [9].

Stimulated emission depletion (STED) nanoscopy, i.e. a targeted superresolution microscopy method, has been used to study the static supramolecular organization of the muscle-type nAChR at the cell surface of fixed mammalian cells [10]. In the present work, we employ single-molecule localization microscopy (SMLM) in the form of stochastic optical reconstruction microscopy (STORM) to image live CHO-K1/A5 cells labelled with a monovalent ligand, fluorescent α-bungarotoxin (BTX). We followed the translational dynamics of the nAChR at the cell surface using single-particle tracking (SPT) methods. The experimental data was subjected to an intertwined combination of analytical and statistical techniques, disclosing a wide distribution of diffusivities, with a predominant mixture of subdiffusive and Brownian regimes. The physical models tested to reveal the source of the anomalous diffusion suggest that coexisting macromolecular self-crowding of the nAChR and obstacles in the form of percolating barriers operate in the time scale of tens of milliseconds to seconds, hampering free nAChR diffusion. We were able to dissect in real time the burst-like assembly and disassembly of the receptor nanoclusters, and their duration, size and density. Moreover, a correlation could be established between the behaviour at the individual trajectory level and the heterogeneous motional behaviour at the ensemble level: we found that the nanoscale dynamics of the single-trajectories dictate the population dynamics at the mesoscopic level, and that cholesterol fine tunes free walk diffusion and the stability of confinement zones.

## RESULTS
**Heterogeneity of nAChR mobility**

Live CHO-K1/A5 cells labelled with fluorescent Alexa Fluor$^{555}$ α-bungarotoxin (BTX) were imaged in STORM buffer at a rate of 100 fps. Regions of interest (10 x 10 µm) were selected from an average of 10 cells for each experimental condition. Upon reconstruction of the STORM images, the validated localizations were found to be distributed nonuniformly across the cell surface in the form of puncta of varying fluorescence intensities (Figure 1a). Supplementary Figure 1 shows the precision of the localizations. A representative example of single-molecule BTX-labelled nAChR trajectories with more than 50 steps (some up to 1600 steps) recorded from ventral surface membrane areas of live CHO-K1/A5 cells is shown in Figure 1b. Visual inspection of the tracks already reveals the heterogeneity of the trajectories: all sets of fluorescent-labelled samples exhibit free walks of variable stretch and regions of apparent confinement in restricted domains. Since the weight of truly immobile particles could affect the outcome of the analysis of the mobile population, this observation called for a distinction to be made between confined but mobile receptors and truly immobile ones, and to exclude the latter.

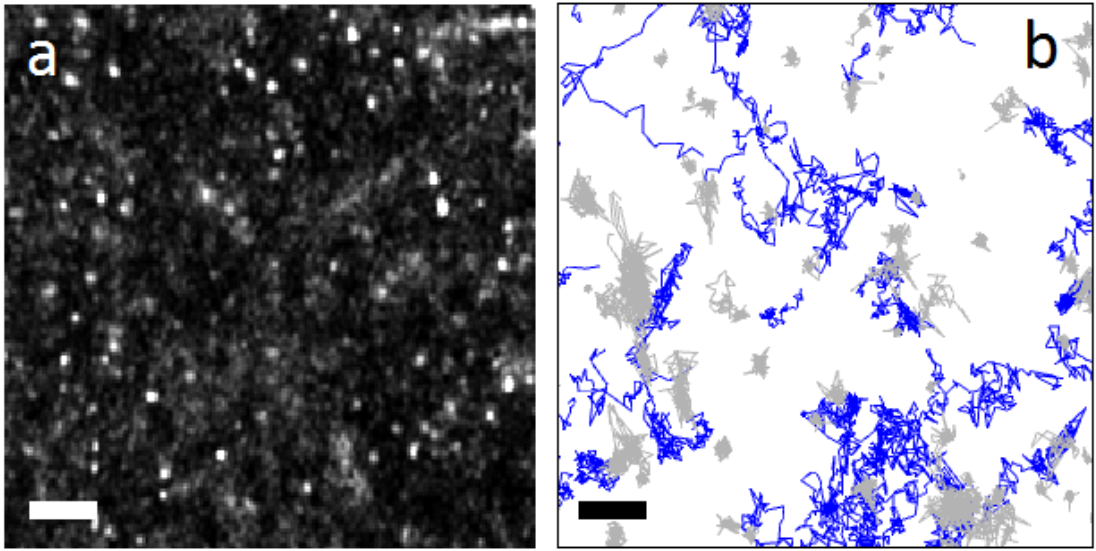

**Figure 1. Raw localizations and nAChR trajectories.** a) A 3.8 x 3.8 µm Gaussian-smoothened projection image of a CHO-K1/A5 cell stained with AlexaFluor[555]-BTX and subjected to cholesterol enrichment (CDx-Chol), showing the identified and validated localizations. b) Mobile (blue) and immobile (grey) nAChR trajectories as defined in ref. [11]. Scale bar: 500 nm.

In order to exclude the immobile molecules from the analysis, we next applied a recently introduced sorting procedure [11] which combines the radius of gyration and the mean displacements of the particles (see Supplementary Figure 2b). The normalized ratio ($\sqrt{\pi/2}\ (R_g/\langle|\Delta r|\rangle)$) obtained from the experiments with *fixed* cells is a constant independent of the localization error or the threshold diffusion coefficient ($D_{th}$) [11]. Using this approach, we calculated the threshold value to exclude immobile molecules in our *live* cell experiments (Suppl. Figure 2c). Threshold values of 1.5-2.1 (95% confidence) were obtained by pooling data from different cells in two independent sets of experiments, and the conservative value of 2.1 was chosen. The pooled mean values of immobile receptors varied between 50 and 65% (Suppl. Figure 2). The immobile nAChR trajectories were excluded from further analysis.

To characterize the heterogeneous diffusional behaviour of the nAChR mobile molecules upon removal of stationary ones, we first analysed the mean-square displacements (MSDs). The MSD curves can be approximated as $\sim t_{lag}^{\beta}$, as explicitly stated in Eq. 4 in Suppl. Material. By linearly fitting the log-log transformed MSD curves we obtained the power-law scaling, i.e. the anomalous exponent β. Figure 2b shows the cumulative density function (CDF) of the anomalous exponent β for the entire (unsorted) populations of control samples and their corresponding curves upon cholesterol depletion or enrichment. The criterion for the choice of β is the same as that adopted in ref. [12], i.e. the use of a non-dimensional parameter. The CDF graphs show that cholesterol enrichment increases diffusivity, albeit to a small extent (β values were 0.92 ± 0.02, 0.96 ± 0.03 and 1.02 ± 0.02 for cholesterol-depletion, control, and cholesterol-enrichment conditions, respectively). The wide span covered by the β exponent, from ca. 0.3 to 1.6, leads us to consider the possibility that the high heterogeneity of the nAChR mobilities could be a consequence of ergodicity breaking.

**nAChR trajectories display weak ergodicity breaking**

To evaluate whether the observed anomalous nature of nAChR diffusion was associated with ergodicity breaking, we compared the time-averaged and ensemble-average MSDs for the whole population of trajectories. The tMSD plots include all the individual trajectories and the fit corresponds to the linear regression to the average of the individual time-averaged trajectories (in log scale), whereas the eMSD distribution results from averaging the square displacement of all trajectories occurring at a given interval (Figure 2a). The span of the tMSD is broad in all cases, indicating that they do not self-average. The values of β are indicated in the two plots, and a comparison of all the experimental conditions is given in Table 1. In ergodic systems, tMSD and eMSD converge to similar values for large numbers of diffusing molecules and long enough times. Here, control samples exhibit weak ergodicity breaking (Table 1). The MSD data provide additional information on the changes induced by cholesterol on nAChR mobility (Suppl. Figure 3), especially in the case of cholesterol-enriched samples, where the ensemble diffusivity was markedly increased (Figure 2b). Figure 2c shows the corresponding generalized diffusion coefficient $K_\beta$ of nAChR single molecules under control and cholesterol-modifying conditions, which also exhibited a broad distribution covering more than two decades ($0.23 \pm 0.04$ $\mu m^2 s^{-\beta}$, $0.26 \pm 0.06$ $\mu m^2 s^{-\beta}$, and $0.43 \pm 0.14$ $\mu m^2 s^{-\beta}$, for cholesterol depletion, control, and cholesterol enrichment conditions, respectively). Highly significant statistical differences were observed between the $K_\beta$ values of cholesterol-enriched and control samples (p<0.0001), mimicking the trends observed with the exponent β.

**Table 1. Ergodicity analysis of nAChR trajectories applied to the entire population of BTX-labelled nAChR trajectories under different cholesterol-modifying conditions***

|      | CDx | Control | CDx-Chol |
|------|-----|---------|----------|
| tMSD | 0.92 ± 0.02 | 0.96 ± 0.03 | 1.02 ± 0.02 |
| eMSD | 0.90 ± 0.02 | 0.83 ± 0.03 | 0.96 ± 0.04 |

*The anomalous exponent β corresponds in both cases to the slope obtained from the linear fitting to the log-log plot of the single molecule average tMSD and the corresponding eMSD, i.e. the average across the ensemble of trajectories as described under Material and Methods in the Supplementary Material. Mean ± 95% confidence intervals are shown.

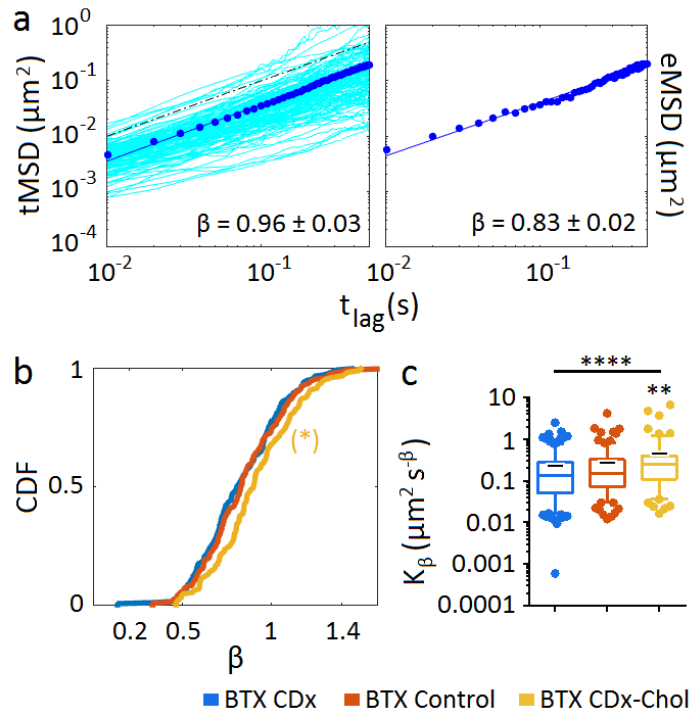

**Figure 2. Time-averaged (tMSD), ensemble-averaged MSD (eMSD), cumulative distribution function of anomalous exponent β and generalized diffusion coefficient $K_\beta$ distribution of the total populations of nAChR trajectories, under control and cholesterol-modifying conditions**. a) The top-left panel is the log-log plot of the tMSD corresponding to control nAChRs including all the individual trajectories having a goodness of fit better than 0.9 (see Suppl. Material). The dots correspond to the average tMSD and the solid line is the fit to the log-log transformed data. The exponent β values ± 95% confidence interval are indicated in each case. The dashed line is a visual aid that scales linearly with the log of time. The top-right panel shows the linear fitting (blue full line) to the log-log transformed eMSD (solid points). The fits scale as $\beta \log(t_{lag})$, rendering the power (anomalous) exponent β. b) Cumulative density distribution function (CDF) of the anomalous exponent β. c) Effect of cholesterol depletion (CDx) or enrichment (CDx-Chol) on the generalized diffusion coefficient, $K_\beta$. Whiskers in box plots correspond to 95% confidence intervals; the limits indicate 75% confidence intervals; the black – symbols indicate the mean and the horizontal lines the median in each case. The dots outside the confidence intervals are outliers. Statistics: p<0.05 (*), p<0.01 (**) and p<0.0001 (****).

**Underlying physical mechanisms: Escape (waiting) time and turning angle probability analyses**

To investigate the possible underlying physical substrate of the observed nAChR subdiffusive behaviour, we resorted to analysis of the distribution of two variables: the probabilities of the escape times and of the turning angles.

Continuous time random walk (CTRW) is a non-ergodic anomalous diffusion model consisting of random walks with transient immobilization, with a dwell-time probability density that scales as $\sim t^{-(\beta+1)}$. If $\beta \leq 1$, the mean dwell-time diverges, and the experimental time window cannot reach the characteristic time of the system, leading to ergodicity breaking. The escape (waiting) time is the interval during which a trajectory remains within a given radius $R_{TH}$ and can be used to test whether the CTRW model explains the experimental data. If the long-time dynamics of trajectories is dominated by the immobilization events, then the waiting time distribution should not depend on the escape radii. We quantified the duration of the events in which the molecules' trajectories remained within circular areas of increasing radii $R_{TH}$ [13, 14]. The distribution of waiting times showed a clear dependence on escape radius between ~20 to 250 nm (Figure 3a), strongly suggesting that the ergodicity breaking observed under control conditions (see also Table 1) cannot be satisfactorily accounted for by the CTRW model.

We next analysed the directional changes in individual nAChR trajectories using the relative angles distended by the molecules along their walk, a parameter which can help distinguish among different types of subdiffusive mechanisms [15, 16]. The turning angle distribution tests for correlations in the particle displacements (see graphical explanation in Suppl. Figure 5). When applied to the total population, the probability of the turning angles was found to increase gradually from 60-90° onwards to peak at $\theta$ = 180° for *all* experimental conditions (Figure 3b), a clear indication of anticorrelated steps. A time-lag dependence was apparent in the directional changes of the control and cholesterol-depleted conditions, suggesting diffusion in a meshwork, a condition compatible with permeable fence models [15].

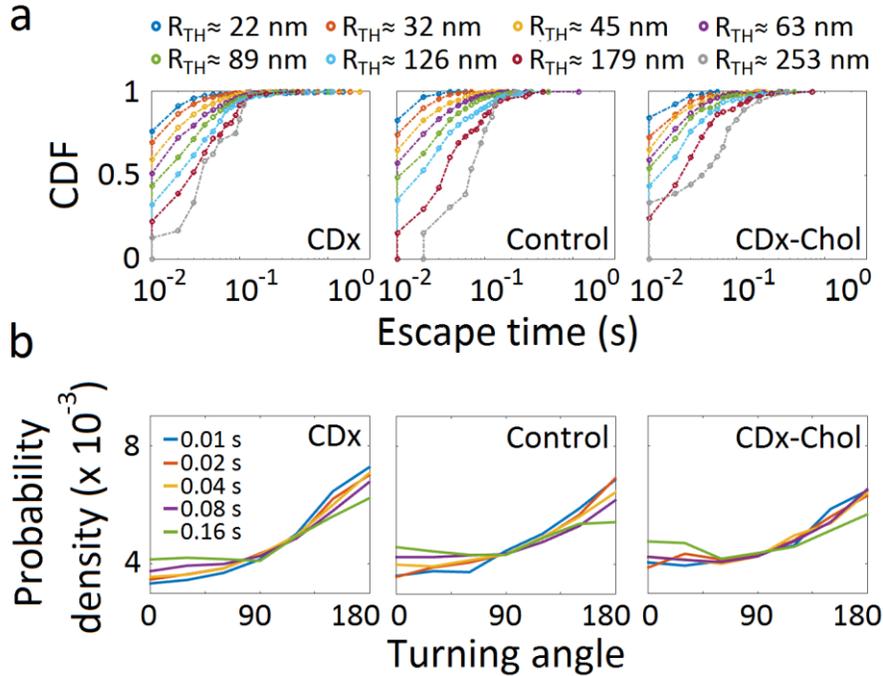

**Figure 3. Escape (waiting) time distributions and turning angle probability densities of nAChRs**
a) Cumulative density (distribution) function of the escape (waiting) times, corresponding to increasing radii $R_{TH}$ from 22 nm to 253 nm. b) Probability densities for the colour-coded $t_{lag}$ of increasing durations (10 to 160 ms). The probability density is normalized such that the integral of the curve is equal to unity.

**Classification of trajectories into diffusivity-based subgroups**

We next separated nAChR trajectories according to their diffusivities. Based on the power exponent β, trajectories can be grouped into subdiffusive I ($\beta < 0.5$), subdiffusive II ($0.5 < \beta \leq 0.7$), subdiffusive III ($0.7 < \beta \leq 0.9$), Brownian ($0.9 < \beta \leq 1.1$), and superdiffusive ($\beta > 1.1$) subpopulations. Using this classification, several properties of the different motional motifs could be identified, which were otherwise hidden in the unsorted population (Figure 3). Whereas the escape (waiting) time of the subpopulations is like that of the ensemble population of trajectories (Figure 4a), the turning angle probability distribution shows distinct features: it increases from 60-90° onwards to peak at $\theta = 180°$ for subdiffusive subpopulations, with a progressively decreasing slope from the subdiffusive I to the subdiffusive III subpopulations; in contrast, the Brownian trajectories exhibit a nearly flat and uniform turning angle distribution with anticorrelated steps at short time lags and correlated steps at long time lags (Figure 4b). The same tendency is observed in the superdiffusive subpopulation, which exhibits a more pronounced step correlation (peak at $\theta = 0°$) at long time lags (40 ms to 160 ms). Similar trends in angular dependence are observed under cholesterol-modifying conditions (see Suppl. Figure 6).

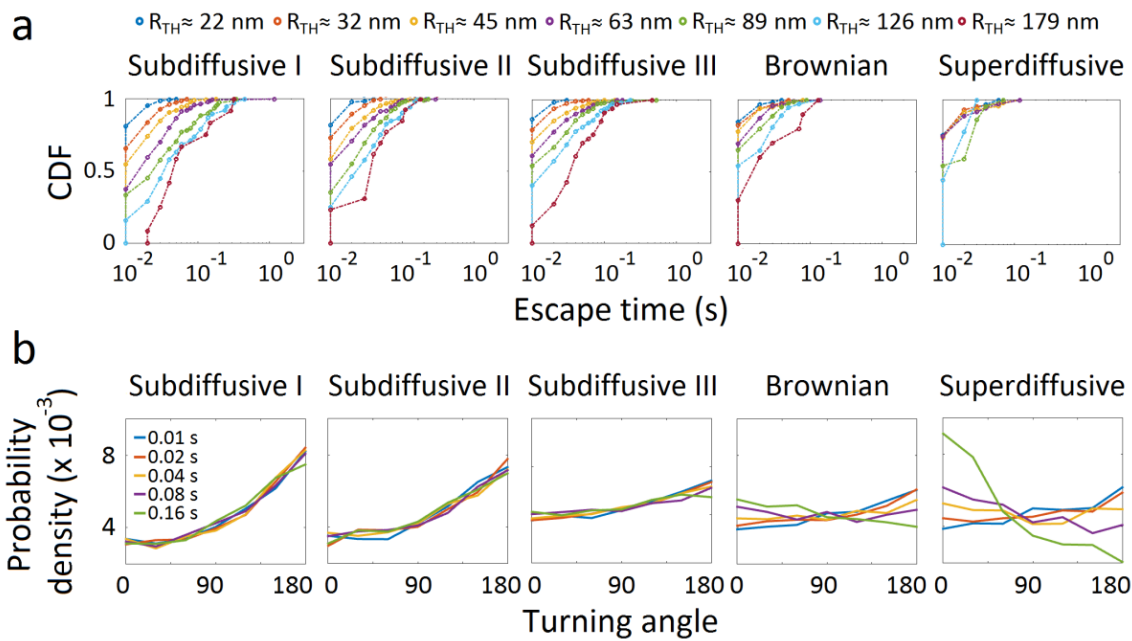

**Figure 4. Analysis of diffusivity-based trajectories.** a) Cumulative density function (CDF) of escape (waiting) times corresponding to increasing radii $R_{TH}$ for the diffusivity-based subpopulations under control conditions. b) Turning angle probability density distribution of each subpopulation (control condition). Turning angles correspond to increasing time lags indicated by the colour code, from 10 ms (light blue) to 160 ms (green). The probability density is normalized such that the integral of the curve is equal to unity.

**Microscopic heterogeneity: transient confinement nanodomains within individual trajectories and differences between subdiffusive states**

We next focused on the behaviour of individual trajectories. To this end, we applied a series of recently introduced analytical tools [17] to scrutinize in step-by-step detail the single-molecule tracks. One such tool (recurrence analysis) examines the recurrence of individual trajectories, i.e. whether a molecule stops at the same location that it has previously visited at the membrane surface. This analysis discriminates between different dynamics and characterizes the shape of the confinement areas, rendering the signature motional behaviour at the single-molecule level. Remarkably, essentially all trajectories, independently of their diffusional modality, exhibit an intermittent alternation between unconfined free-diffusing portions and sections of the track in which the molecule is confined for varying intervals within small (nanometre-sized) areas (Figure 5a). The ellipsoidal areas covered by the confinement sojourns have an average major semi-axis of ca. 40 nm (39 ± 2 nm, 45 ± 3 nm and 46 ± 6 nm for cholesterol-depletion, control, and cholesterol-enrichment conditions, respectively) and an eccentricity of ~0.6. The ellipsoidal surfaces are equivalent to circular areas with radii of 32 ± 2 nm, 36 ± 2 nm and 37 ± 4 nm for cholesterol depletion, control, and cholesterol enrichment conditions, respectively. The longest residence times in the confinement sojourns (Figure 5b) correspond to the subdiffusive subpopulation I and II, followed by the subdiffusive III, Brownian and superdiffusive groups, in that order. Changes in cholesterol levels modulate the lifetime of the confinement sojourns, more markedly for the less diffusive subpopulations (Figures 5c-d). Also, the confinement areas diminish in size upon cholesterol depletion (Figure 5e). The proportion of the trajectories spent in the confined state is ~30%, changing upon cholesterol modification (Suppl. Fig. 9b). Furthermore, marked differences are observed in the % of the confined state per single-molecule track between the different diffusivity-based subpopulations, especially for the subdiffusive ones (Suppl. Figure 9a), an observation that led us to analyse them separately. Analogously, whereas the cumulative distribution function β corresponding to the free portions of the unsorted, averaged population varied from Brownian to nearly superdiffusive (β = 1.07 ± 0.08, 1.10 ± 0.08 and 1.10 ± 0.08 for cholesterol-depletion, control, and cholesterol-enrichment conditions, respectively), the subdiffusive behaviour is only apparent in the confined portions (Figure 6b), suggesting that the confinement sojourns in nanodomains are the ones that determine the predominant subdiffusive behaviour of the nAChR population.

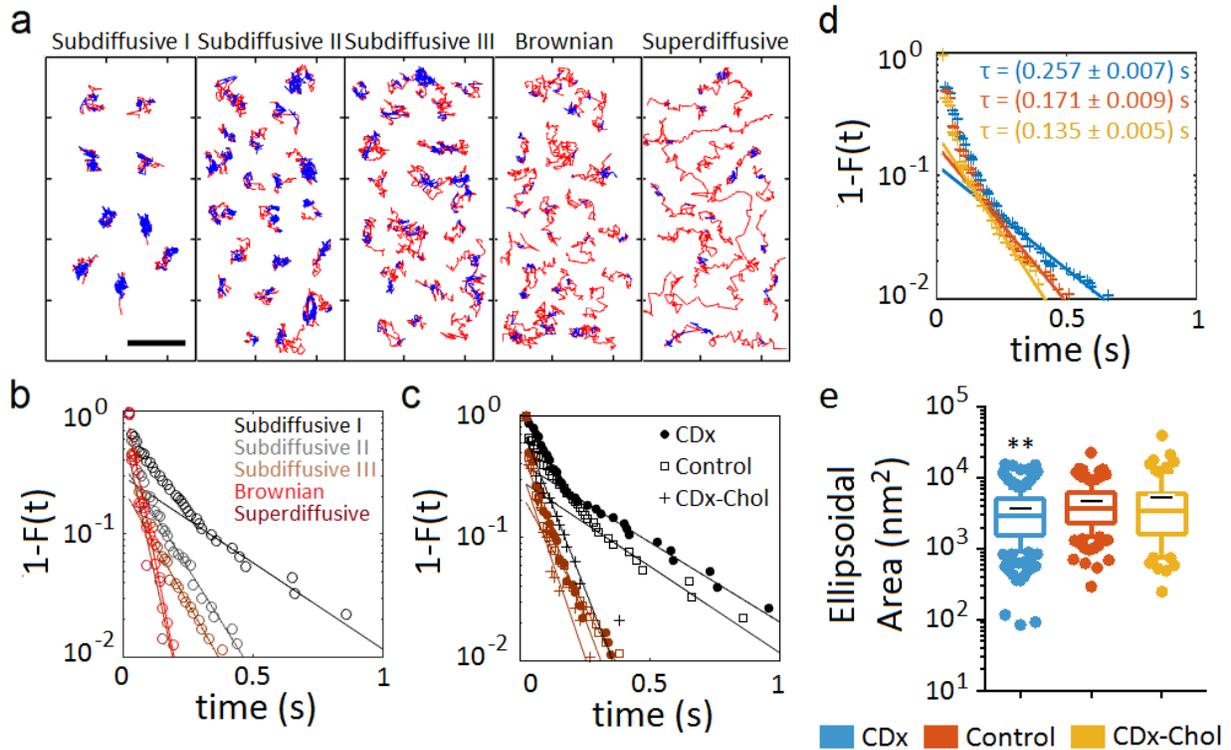

**Figure 5. Microscopic heterogeneity: transient confinement in nanodomains within individual trajectories and differences between subdiffusive states.** a) Recurrence analysis of individual trajectories [18] identifies the unconfined diffusing portion of each individual trajectory (red) and transient confinement sojourns (blue) within the trajectories under control conditions. Notice the decreasing immobilization (confinement) from the subdiffusive I to the superdiffusive subpopulation. Bar: 1 µm. b) Complementary cumulative distribution function of the residence times in the confined state of each subpopulation (control conditions). The tails of the curves were fitted with exponential decay functions, from which the decay times were obtained (listed in Suppl. Table 2). c) The same as in (b) but showing the effect of cholesterol modification on the subdiffusive I and III subpopulation (same colour code as in (b)). d) The same as in (c), but for the total population of nAChRs after cholesterol enrichment and depletion. e) Ellipsoidal area of the trajectories´ portion in the confined state, obtained by fitting the confinement domain in the trajectory with an elliptic surface. Whiskers in box plots correspond to 95% confidence intervals; the limits indicate 75% confidence intervals; the black – symbols indicate the mean and the horizontal lines the median in each case. The dots outside the confidence intervals are outliers. (**), $p<0.01$.

The ability to dissect individual trajectories into confined and free regions (Figure 5) additionally allowed us to apply turning angle analysis to the two sections separately. The confined regions show a clear anticorrelation (peak at $\theta = 180°$) and time invariance (i.e. independent of time lag) for all experimental conditions (Figure 6a). In contrast, the free (unconfined) portions appear to conform to the behaviour typically observed for Brownian diffusion, i.e. an angle-independent, nearly flat uniform distribution, but exhibit time lag dependence, Brownian-like flatness at short times and are positively correlated, with a peak at $\theta = 0°$ at long times. Clear statistical differences are observed between the anomalous exponent β of free and confined trajectories (Fig. 6b; p< 0.01 and p< 0.0001 for control and cholesterol-enriched conditions, respectively).

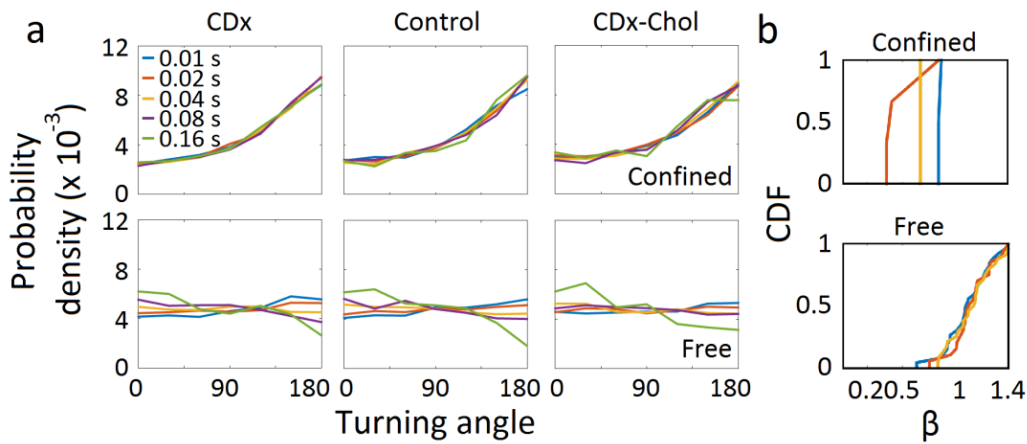

**Figure 6. Dynamic characterization of the transiently confined and unconfined portions of the individual trajectories in control and under cholesterol-modifying conditions.** a) Turning angle distribution corresponding to increasing lag times (indicated in the top left panel), from 1 frame (10 ms, blue) to 16 frames (160 ms, green) for the confined portions and the unconfined, free portions of the trajectories having at least 50 steps. The probability density is normalized such that the integral of the curve is equal to unity. b) Distribution of the anomalous exponent β for free and confined portions in the ensemble, total population of nAChRs under cholesterol-depletion and -enrichment.

**Disclosure of nAChR nanocluster formation/disassembly dynamics in real time**

We next employed the qSR algorithms developed by Cisse and coworkers [19] to follow nAChR nanocluster dynamics in real time. As shown in Figure 7a, bursts of activity in the form of vertical spikes (ascending coloured cumulative count trace), and periods of "inactivity" (grey horizontal lines in the cumulative count trace), conform the signature of cluster formation/dissociation [20, 21]. The spatial arrangement of the single-molecule localizations belonging to each dynamic nanocluster in the confinement regions of the traces is shown in Figure 7b. Cholesterol modification induces marked statistical differences in the nanocluster metrics: Nanoclusters exhibited RMS radii of 107 ± 4 nm, 149 ± 13nm and 149 ± 13 nm for cholesterol depletion, control, and cholesterol enrichment conditions, respectively; lifetimes in the order of seconds (2.45 ± 0.14 s, 3.90 ± 0.77 s and 2.96 ± 0.32 s for cholesterol depletion, control, and cholesterol enrichment conditions, respectively); and tens of particles per cluster (16 ± 1, 30 ± 8 and 20 ± 3 particles for cholesterol depletion, control, and cholesterol enrichment conditions, respectively) (Figure 7c). No statistical differences were observed between control and cholesterol-modified samples in the intra-burst ("dark") periods lasting ~0.17 s.

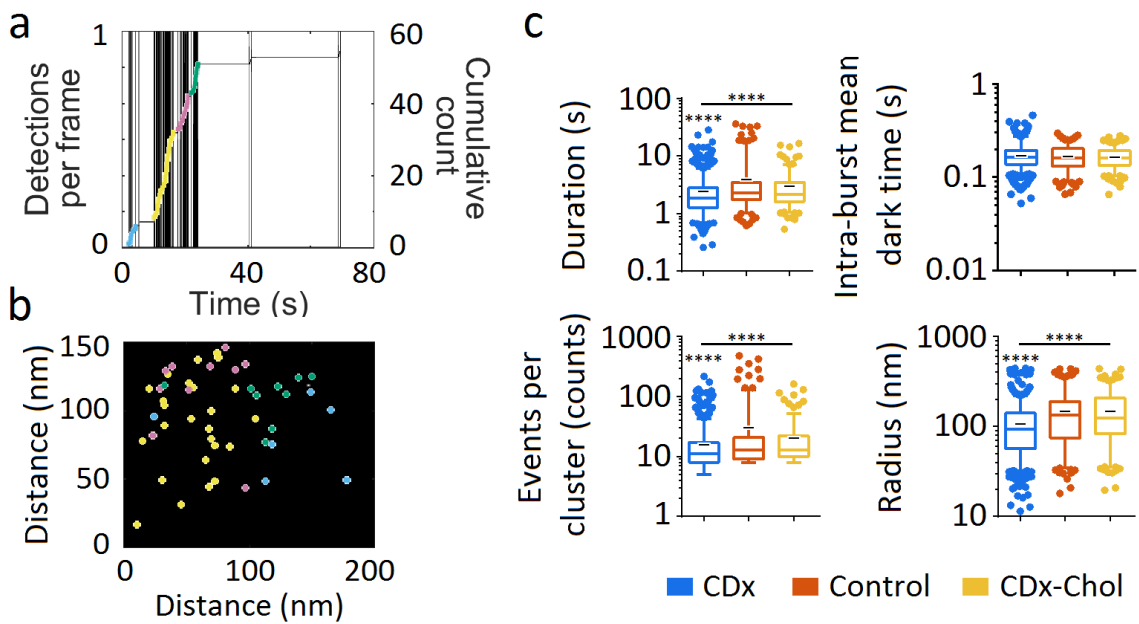

**Figure 7. Real time follow-up of nAChR dynamic nanoclusters under cholesterol depletion and enrichment**. a) Using the live-cell single-molecule localizations validated with ThunderSTORM as the input data, the qSR software was applied to generate the time-dependent plot: The bursts of individual spikes identify the formation of dynamic nanoclusters; the cumulative count traces (ascending continuous coloured line) superimposed on the spikes delimit each (nano)cluster in a different colour. The horizontal plateau sections of the trace correspond to the nanocluster breakages[20]. (b) A 150x200 nm region showing single-molecule cluster localizations (corresponding to the cumulative traces shown with the same colour in (a)). c) nAChR nanocluster metrics (log scale): duration, intra-burst mean dark time, i.e. periods without localizations within a cluster, events per cluster, and RMS radii of clusters [21] under control and cholesterol modifying conditions. Whiskers in box plots correspond to 95% confidence intervals; the limits indicate 75% confidence intervals; the black negative symbols indicate the mean and the horizontal lines the median in each case. The dots outside the confidence intervals are outliers. Statistics: (****), $p<0.0001$.

## Discussion

The ultimate aim of this work is to investigate how variations in synapse cholesterol content influence nAChR mobility. We have chosen a model mammalian heterologous clonal cell line produced in our laboratory which robustly expresses adult-type nAChR [22] to test in a straightforward manner a focused set of objectives of this aim, technically inaccessible in the intact synapse, using state of the art high-density single molecule tracking, single-molecule localization-based superresolution microscopy, and computational analytical tools.

**Heterogeneous behaviour of the ensemble population of nAChR trajectories**

The first general observation stemming from our multi-pronged analyses concerns the complexity and heterogeneity of nAChR dynamics. The cell-surface membrane of mammalian cells is a complex environment from both structural and functional viewpoints. It is in fact surprising that some membrane proteins manage to diffuse by thermally-driven, viscosity-dependent, uncorrelated Brownian motion. Deviations from Brownian motion, i.e. anomalous diffusion, in such a crowded and heterogeneous milieu is more often than not the rule [23-25], as has also been observed with the nAChR [26-28].

Recently He and coworkers [29] followed the mobility of quantum-dot labelled nAChRs in *Xenopus* muscle cells in culture. The nAChR did not obey the Gaussian statistics characteristic of Brownian diffusion. They interpreted these observations as resulting from dynamic heterogeneity, and proposed a variant of Kusumi´s picket-fence model [30, 31] which they termed "dynamic picket-fence model". Kusumi´s original model purports the existence of "corrals" formed by sub-membrane actin filament network "fences" converging onto "pickets" of actin-binding proteins. Pickets and fences act as percolation barriers that diffusing proteins must "hop over" to dodge the corrals. In the present experiments, we observe confinement areas with a normalized radius of ~36 nm in control samples; actin corral immobilization operates on membrane areas in this spatial scale [31-33]. He and coworkers [29] also argued that immobile nAChR tracks may contribute to the non-ergodic MSDs, but in the present work we have specifically addressed and discarded this possibility using recently established criteria [11] to exclude immobile trajectories.

**Physical mechanisms behind the anomalous nAChR dynamics**

Ergodicity lies at the core of statistical mechanics; the convergence of the temporal and the ensemble averages is known as the ergodic hypothesis (see review in ref. [34]). Diffusion processes that deviate from Brownian motion are considered anomalous, and their propagators may or may not be Gaussian [14, 35]. However, non-ergodic systems can be conformed to nearly Gaussian, probably due to the probe only sampling a local region during the time scale of the observation [36]. This is particularly important for some models having slow convergence, such as the obstructed diffusion (OD) model with obstacle concentrations close to criticality [34]. Our experiments cover a limited temporal window, within which the comparison of the time-averaged MSD with the ensemble-averaged MSD revealed the presence of both ergodic and non-ergodic behaviours at the ensemble level (Figure 2 and Table 1). We therefore attempted to discriminate between plausible physical models accounting for such complexity.

The distribution of relative angles of motion between successive time intervals of a trajectory provides information beyond that afforded by MSD analyses [15, 16]. A preferred angle of 0° indicates that the moving particle travels in the same direction as the previous step; in contrast, a preferred angle $\theta$ = 180° specifies that the molecule walks in the opposite direction to the previous step. The latter appears to be the case with the trajectories of the total population (Figure 3) as well as with all subdiffusive trajectories (Figure 4 and Suppl. Figure 6), indicating that nAChR molecules turn back rather than continue their motion in a defined direction, a fingerprint of subdiffusive random walks with anticorrelated increments [15, 34]. Could this be accounted for by viscoelastic properties of the membrane or could it be the result of interactions with physical obstacles? Turning angle analysis allowed us to discriminate between these two scenarios. The fractional Brownian motion (fBM) model (a generalization of Brownian motion with power-law correlated displacements and a Gaussian propagator) describes diffusion in viscoelastic fluids, whereas the OD model refers to particles hindered by near-immobile obstacles. Unlike Brownian motion, in the fBM model particles revisit previously visited locations [37] and exhibit a time-dependent diffusion coefficient [17, 34, 38]. Krapf and coworkers [15] have pointed out that although both the fBM and the OD models display turning angle distributions peaking at $\theta$ = 180°, fBM exhibits a gradual increase between 45° to 135° and a plateau at higher angles, whereas the OD model shows a continuous increment between 90° and 180°. The subdiffusive nAChR trajectories exhibit the latter signature (Figures 3-4 and Suppl. Figure 6), suggesting the fBM model should be discarded in favour of the OD model and indicating the presence of obstacles impeding or delaying nAChR motion. The anticorrelated steps are time lag-dependent, a feature consistent with diffusion in a meshwork [15] and the picket-fence model (see recent update in ref.[39]). In comparison with the numerical simulations of ref. [15] the steepness of the increase in the subdiffusive type I and II subpopulations suggests obstacle concentrations close to criticality, and the declining slope from the subdiffusive I towards the Brownian subpopulation submits to either i) a decreasing gradient of obstacle concentration [15] or the possibility that ii) the ratio of free/confined regions of the individual trajectories increases accordingly. Characteristic obstacles for membrane protein lateral diffusion are protein self-aggregation, pickets of other immobile or slowly moving proteins, sub-membrane actin corrals, or liquid-ordered (Lo) lipid heterogeneities ("rafts"). Any of these obstacles, or mixtures thereof, add complexity to the mechanism of transient immobilization and can give rise to apparent non-ergodic behaviour [40], particularly because they may act on different spatial and time scales. This may be the case with the nAChR trajectories that exhibit weak ergodicity breaking.

Examination of the duration of the events in which the confined region of the trajectories remained within circular areas of increasing radii $R_{TH}$ [13, 14] provided the opportunity to test the CTRW model [23-25]. In this hypothesis, molecules move randomly, with waiting times also random, with a probability $\tau^{-(\beta+1)}$, and average tMSD linearly scaling with lag time. CTRW is a non-ergodic model of subdiffusion associated with binding to stationary components such as the actin meshwork [14], molecular crowding, protein-protein interactions or lipid domains, all of which may give rise to transient immobilization. Instead, a higher, close-to-criticality obstacle concentration could explain the phenomenon, resulting in turn from e.g. stabilization of the actin cytoskeleton [41] with convergence times beyond the temporal window explored in the present work.

**Microscopic heterogeneity: transient confinement nanodomains interrupt individual trajectories**

Application of a series of analytical tools recently introduced by Krapf and his group to inspect individual trajectories in detail[17] led us to disclose the occurrence of pauses -remarkably in *all* single nAChR trajectories, including Brownian and, less frequently, even superdiffusive trajectories. The latter may result from active processes associated with molecular motors/the submembrane cytoskeleton or have their origin in clustering/de-clustering events like those observed with submicron tracers in micellar solutions. The mobile tracks were interrupted in all cases by sojourns of variable duration, whereby receptors were transiently confined within nm-sized domains of ca. 36 nm. Other SPT studies reported that 20% of nAChR in muscle myoblasts showed restricted diffusion in small domains having similar dimensions [27]. We found that the less diffusive subpopulations spent the longest stopovers in the confined state; the confined sojourns of Brownian trajectories had shorter durations, with superdiffusive trajectories displaying the shortest confined lifetimes (Figure 5, Suppl. Figure 7 and Suppl. Table 2). Thus, receptor dynamics appear to be governed by the probability of encounter and the time spent in confined and free-diffusing regions of the trajectory. Modification of the cholesterol levels changed the characteristic lifetime and the areas of the confined sojourns (Figure 5d-e).

**Modulation of nAChR diffusional dynamics: synergistic effects of cholesterol and receptor self-crowding and trapping in confined regions**

Cholesterol possibly affects nAChR diffusivity through two non-exclusive and complementary mechanisms: i) via the physical state of the bulk bilayer, overwhelmingly determined by the chemical composition of the constituent lipids, and ii) through changes in the abundance and/or size of cholesterol/sphingolipid-rich nanodomains. Mechanism (ii) appears to influence mainly the s-long kinetics of formation and disassembly of nAChR nanoaggregates which we could follow in real time (Figure 7). Additionally, the turning angle behaviour of the nAChR walks with anticorrelated steps (Figures 3-4 and 6), can be satisfactorily accounted for by the occurrence of cholesterol/sphingolipid-rich Lo domains acting as lateral heterogeneities in the plane of the membrane, i.e. obstacles that nAChRs have to circumvent in a permeable fence scenario, in accordance with the notion that heterogeneity of the diffusing bodies' environment has a strong impact on mobility [36]. The possibility that nanometre-sized Lo domains have a role (direct or indirect) in inducing confinement finds independent experimental support. When reconstituted in a sphingomyelin-cholesterol-POPC (1:1:1) model system, purified nAChR from *Torpedo* does not exhibit preference for partitioning in Lo domains [42]. However, inclusion of sphingomyelin molecular species that generate bilayer asymmetry by enriching the sphingolipid content of the outer leaflet appear to favour the inclusion and enrichment of the nAChR in Lo domains [43]. Thus, nAChRs can inhabit/be excluded from Lo domains ("rafts") depending on the composition of the local lipid microenvironment, and this, in turn, facilitates receptor self-aggregation. This is particularly relevant to the function of the cholinergic synapse: in addition to the quite high concentration of cholesterol compared to other membrane lipids in the neuromuscular junction or in the *Torpedo* electromotor synapse [44], one outstanding feature of these two synapses is the extraordinarily high density of nAChR protein [45], a feature shared with the CHO-K1/A5 cell line [10] studied here albeit to a less pronounced degree. Macromolecular crowding in cell membranes is known to produce deviations from Brownian diffusion (reviewed in ref. [46]). In our hypothesis, the cholesterol-rich Lo domains could hinder or reduce diffusion and concentrate sluggish or immobile nAChRs. We excluded immobile receptors from the diffusional analysis, but one must bear in mind that these static macromolecules constitute nonetheless a major proportion of the total population (and as such are included in the clustering analysis). Recently, cholesterol-dependent structures coined "cages" were reported in the CHO line, parent cells of our CHO-K1/A5 line [47]. Cages are larger and longer-lived than the purported size of some "raft" domains and are stabilized within even larger and longer-lived actin corrals. Cholesterol depletion destabilizes these diffusional barriers, as we observe with the nanoclusters in the present work (Figure 7). The confinement nanodomains that we observe (~36 nm radius) are similar to others recently disclosed in static imaging of cholesterol-dependent nanodomains, such as those harbouring dopamine receptors (~35 nm radius [48]), or GPI-anchored protein raft-born nanodomains (~40 nm [39]).

The question arose as to whether the different diffusional regimes are due to different concentrations of obstacles or to the relative time spent in confined/free zones. When we dissected the individual trajectories into confined and free regions, it became clear that the turning angles are essentially the same in each region, indicating that the nAChR in confinement regions encounters similarly high obstacle concentrations -close to criticality- or a similar probability of permeation under all experimental conditions (Figure 6), affecting *all* diffusive regimes (Suppl. Figure 8) in marked contrast to the behaviour of the free diffusing regions (Figure 6 and Suppl. Figure 8). This is a key observation: the time spent in confined sojourns relative to that in free walks determines the resulting motional regime and the breadth of the macroscopic heterogeneity in the ensemble population.

In addition to the transient and reversible nanoscale confinement sojourns within single nAChR trajectories, lasting tens of milliseconds along the individual walks, we disclosed another dynamic process, occurring in the time window of seconds and of a greater spatial extension. This much slower dynamic process likely corresponds to the assembly/residence/disassembly of nanoclusters within confinement areas. We observed statistically significant differences in the RMS radius and number of events in these nm-sized nanoclusters, as well as in the lifetime of the nanocluster assembly period, following changes in cholesterol content (Figure 7). Thus, the alternating free diffusion + confined diffusion at the single-molecule level is reflected at the mesoscopic level in the "social" behaviour of the receptor, i.e. its tendency to reversibly self-associate in a cholesterol-dependent manner in supramolecular aggregates at the cell surface. The signature link between the two processes is their cholesterol dependence, modulating diffusion in the bulk bilayer and stabilizing nanodomains in the confinement regions.

**References**
1. Barrantes, F.J. Cholesterol effects on nicotinic acetylcholine receptor. *J Neurochem* **103 Suppl 1**, 72-80 (2007).
2. Barrantes, F.J. Phylogenetic conservation of protein-lipid motifs in pentameric ligand-gated ion channels. *Biochim Biophys Acta* **1848**, 1796-1805 (2015).
3. Barrantes, F.J. & Fantini, J. From hopanoids to cholesterol: Molecular clocks of pentameric ligand-gated ion channels. *Progress in lipid research* **63**, 1-13 (2016).
4. Mantipragada, S.B. *et al.* Lipid-protein interactions and effect of local anesthetics in acetylcholine receptor-rich membranes from Torpedo marmorata electric organ. *Biochemistry* **42**, 9167-9175 (2003).
5. Barrantes, F.J. Structural basis for lipid modulation of nicotinic acetylcholine receptor function. *Brain Res.Brain Res.Rev.* **47**, 71-95 (2004).
6. Bruses, J.L., Chauvet, N. & Rutishauser, U. Membrane lipid rafts are necessary for the maintenance of the (alpha)7 nicotinic acetylcholine receptor in somatic spines of ciliary neurons. *J Neurosci.* **21**, 504-512 (2001).
7. Campagna, J.A. & Fallon, J. Lipid rafts are involved in C95 (4,8) agrin fragment-induced acetylcholine receptor clustering. *Neuroscience* **138**, 123-132 (2006).
8. Marchand, S., Devillers-Thiery, A., Pons, S., Changeux, J.P. & Cartaud, J. Rapsyn escorts the nicotinic acetylcholine receptor along the exocytic pathway via association with lipid rafts. *The Journal of neuroscience : the official journal of the Society for Neuroscience* **22**, 8891-8901 (2002).
9. Borroni, V. & Barrantes, F.J. Cholesterol modulates the rate and mechanism of acetylcholine receptor internalization. *J Biol Chem* **286**, 17122-17132 (2011).
10. Kellner, R.R., Baier, C.J., Willig, K.I., Hell, S.W. & Barrantes, F.J. Nanoscale organization of nicotinic acetylcholine receptors revealed by stimulated emission depletion microscopy. *Neuroscience* **144**, 135-143 (2007).
11. Golan, Y. & Sherman, E. Resolving mixed mechanisms of protein subdiffusion at the T cell plasma membrane. *Nature communications* **8**, 15851 (2017).

12. Weron, A. *et al.* Ergodicity breaking on the neuronal surface emerges from random switching between diffusive states. *Scientific reports* **7**, 5404 (2017).
13. Manzo, C. *et al.* Weak ergodicity breaking of receptor motion in living cells stemming from random diffusivity. *Physical Review X* **5**, 011021 (2015).
14. Weigel, A.V., Simon, B., Tamkun, M.M. & Krapf, D. Ergodic and nonergodic processes coexist in the plasma membrane as observed by single-molecule tracking. *Proc Natl Acad Sci U S A* **108**, 6438-6443 (2011).
15. Sadegh, S., Higgins, J.L., Mannion, P.C., Tamkun, M.M. & Krapf, D. Plasma Membrane is Compartmentalized by a Self-Similar Cortical Actin Meshwork. *Physical Review X* **7**, 011031 (2017).
16. Burov, S. *et al.* Distribution of directional change as a signature of complex dynamics. *Proc Natl Acad Sci U S A* **110**, 19689-19694 (2013).
17. Sikora, G., Burnecki, K. & Wylomanska, A. Mean-squared-displacement statistical test for fractional Brownian motion. *Phys Rev E* **95**, 032110 (2017).
18. Sikora, G. *et al.* Elucidating distinct ion channel populations on the surface of hippocampal neuros via single-particle tracking recurrence analysis. *arXiv:1708.02876* (2017).
19. Andrews, J.O. *et al.* qSR: a quantitative super-resolution analysis tool reveals the cell-cycle dependent organization of RNA Polymerase I in live human cells. *Scientific reports* **8**, 7424 (2018).
20. Cisse, II *et al.* Real-time dynamics of RNA polymerase II clustering in live human cells. *Science* **341**, 664-667 (2013).
21. Andrews, J.O. *et al.* qSR: A software for quantitative analysis of single molecule and super-resolution data. *bioRxiv* (2017).
22. Roccamo, A.M. *et al.* Cells defective in sphingolipids biosynthesis express low amounts of muscle nicotinic acetylcholine receptor. *The European journal of neuroscience* **11**, 1615-1623 (1999).
23. Saxton, M.J. Single-particle tracking: the distribution of diffusion coefficients. *Biophys.J.* **72**, 1744-1753 (1997).
24. Saxton, M.J. & Jacobson, K. Single-particle tracking: applications to membrane dynamics. *Annual review of biophysics and biomolecular structure* **26**, 373-399 (1997).
25. Jeon, J.H. *et al.* In vivo anomalous diffusion and weak ergodicity breaking of lipid granules. *Phys Rev Lett* **106** (2011).
26. Baier, C.J., Gallegos, C.E., Levi, V. & Barrantes, F.J. Cholesterol modulation of nicotinic acetylcholine receptor surface mobility. *European biophysics journal : EBJ* **39**, 213-227 (2010).
27. Piguet, J., Schreiter, C., Segura, J.M., Vogel, H. & Hovius, R. Acetylcholine receptor organization in membrane domains in muscle cells: evidence for rapsyn-independent and rapsyn-dependent mechanisms. *J Biol Chem* **286**, 363-369 (2011).
28. Almarza, G., Sanchez, F. & Barrantes, F.J. Transient cholesterol effects on nicotinic acetylcholine receptor cell-surface mobility. *PloS one* **9**, e100346 (2014).
29. He, W., Song, H. & Su, Y. Dynamic heterogeneity and non-Gaussian statistics for acetylcholine receptors on live cell membrane. **7**, 11701 (2016).
30. Fujiwara, T., Ritchie, K., Murakoshi, H., Jacobson, K. & Kusumi, A. Phospholipids undergo hop diffusion in compartmentalized cell membrane. *J Cell Biol.* **157**, 1071-1081 (2002).
31. Fujiwara, T.K. *et al.* Confined diffusion of transmembrane proteins and lipids induced by the same actin meshwork lining the plasma membrane. *Mol Biol Cell* **27**, 1101-1119 (2016).
32. Kalay, Z., Fujiwara, T.K., Otaka, A. & Kusumi, A. Lateral diffusion in a discrete fluid membrane with immobile particles. *Physical review. E, Statistical, nonlinear, and soft matter physics* **89**, 022724 (2014).

33. Goswami, D. *et al.* Nanoclusters of GPI-Anchored Proteins Are Formed by Cortical Actin-Driven Activity. *Cell* **135**, 1085-1097 (2008).
34. Krapf, D. Mechanisms underlying anomalous diffusion in the plasma membrane. *Current topics in membranes* **75**, 167-207 (2015).
35. Metzler, R. Gaussianity Fair: The Riddle of Anomalous yet Non-Gaussian Diffusion. *Biophysical Journal* **112**, 413-415 (2017).
36. Munder, M.C. *et al.* A pH-driven transition of the cytoplasm from a fluid- to a solid-like state promotes entry into dormancy. *e-Life* **5** (2016).
37. Metzler, R., Jeon, J.H., Cherstvy, A.G. & Barkai, E. Anomalous diffusion models and their properties: non-stationarity, non-ergodicity, and ageing at the centenary of single particle tracking. *Physical chemistry chemical physics : PCCP* **16**, 24128-24164 (2014).
38. Tabei, S.M. *et al.* Intracellular transport of insulin granules is a subordinated random walk. *Proc Natl Acad Sci U S A* **110**, 4911-4916 (2013).
39. Nemoto, Y.L. *et al.* Dynamic Meso-Scale Anchorage of GPI-Anchored Receptors in the Plasma Membrane: Prion Protein vs. Thy1. *Cell Biochem Biophys* **75**, 399-412 (2017).
40. Mardoukhi, Y., Jeon, J.H. & Metzler, R. Geometry controlled anomalous diffusion in random fractal geometries: looking beyond the infinite cluster. *Physical chemistry chemical physics : PCCP* **17**, 30134-30147 (2015).
41. Kwik, J. *et al.* Membrane cholesterol, lateral mobility, and the phosphatidylinositol 4,5-bisphosphate-dependent organization of cell actin. *Proc Natl Acad Sci U S A* **100**, 13964-13969 (2003).
42. Bermudez, V., Antollini, S.S., Fernandez Nievas, G.A., Aveldano, M.I. & Barrantes, F.J. Partition profile of the nicotinic acetylcholine receptor in lipid domains upon reconstitution. *J Lipid Res* **51**, 2629-2641 (2010).
43. Perillo, V.L., Penalva, D.A., Vitale, A.J., Barrantes, F.J. & Antollini, S.S. Transbilayer asymmetry and sphingomyelin composition modulate the preferential membrane partitioning of the nicotinic acetylcholine receptor in Lo domains. *Arch Biochem Biophys* **591**, 76-86 (2016).
44. Rotstein, N.P., Arias, H.R., Barrantes, F.J. & Aveldano, M.I. Composition of lipids in elasmobranch electric organ and acetylcholine receptor membranes. *J Neurochem* **49**, 1333-1340 (1987).
45. Barrantes, F.J. The lipid environment of the nicotinic acetylcholine receptor in native and reconstituted membranes. *Crit Rev.Biochem.Mol.Biol.* **24**, 437-478 (1989).
46. Metzler, R., Jeon, J.H. & Cherstvy, A.G. Non-Brownian diffusion in lipid membranes: Experiments and simulations. *Biochim Biophys Acta* **1858**, 2451-2467 (2016).
47. Goiko, M., de Bruyn, J.R. & Heit, B. Short-Lived Cages Restrict Protein Diffusion in the Plasma Membrane. *Scientific reports* **6**, 34987 (2016).
48. Rahbek-Clemmensen, T. *et al.* Super-resolution microscopy reveals functional organization of dopamine transporters into cholesterol and neuronal activity-dependent nanodomains. *Nature communications* **8**, 740 (2017).

**Acknowledgments**

Experimental work was supported by grants PIP No. N° 112-201101-01023 and 5205/15 from the National Scientific and Technical Research Council of Argentina (CONICET) and PICT No. 2015-2654/PICT 2015 IB from the Ministry of Science, Technology and Productive Innovation of Argentina to F.J.B. We are grateful to Prof. Stefan Hell and his group, and Dr. Volker Westphal in particular, for help in the construction of the superresolution setup in Argentina.

**Author contributions**

F.J.B. designed and performed research; A.M. and P.C. analyzed the data; F.J.B. wrote the paper, with major contributions from A.M.

# Cholesterol modulates acetylcholine receptor diffusion by tuning confinement sojourns and nanocluster stability

**Alejo Mosqueira, Pablo A. Camino and Francisco J. Barrantes**

Laboratory of Molecular Neurobiology, Institute for Biomedical Research,

UCA–CONICET, Av. Alicia Moreau de Justo 1600, C1107AFF Buenos Aires, Argentina.

## Supplementary Material

**Materials**

Methyl-β-cyclodextrin (CDx), catalase, glucose oxidase, β-mercaptoethanol, polyvinylalcohol (PVA, 25,000 MW, prod. No. 184632) and the monoclonal antibody mAb35 (product M-217) against the extracellular moiety of the nAChR α-subunit were purchased from Sigma Chem. Co. (St. Louis, MO). Native, unlabelled α-bungarotoxin (BTX) and Alexa Fluor-labelled BTX (Alexa Fluor[555]-BTX), Alexa Fluor[555]- or Texas Red-labelled anti-IgG secondary antibodies were purchased from Invitrogen Argentina.

**Cell culture**

CHO-K1/A5 cells were grown in Ham's F12 medium supplemented with 10% foetal bovine serum for 2-3 days at 37°C (ref. [1]) before experiments.

**Acute cyclodextrin-mediated cholesterol depletion/enrichment of cultured cells**

Acute cholesterol depletion was carried out prior to fluorescent labelling by treating CHO-K1/A5 cells with 10-15 mM CDx or CDx-cholesterol complexes in Medium 1 ("M1": 140 mM NaCl, 1 mM $CaCl_2$, 1 mM $MgCl_2$ and 5 mM KCl in 20 mM HEPES buffer, pH 7.4) essentially as in refs.[2, 3]. Samples were taken at 20 min from culture dishes incubated at 37°C in the presence or absence of the cholesterol-modifying chemical.

**Fluorescence nanoscopy setup**

The optical nanoscope used, constructed in our laboratory, operates in the stochastic optical reconstruction microscopy (STORM)[4] /ground state depletion microscopy followed by individual molecule return (GSDIM)[5] modes. A 532 nm pumped solid state DPSS 300 mW laser (HB-Laserkomponenten GmbH, Schwäbisch Gmünd, Germany) was used as the excitation source, and delivered through a 0.65 FCP multiwavelength, polarization-maintaining fibre from Point Source, U.K. The output beam was passed through a dichroic filter (LLF532/10x Brightline, Semrock, Rochester, NY). The laser power was set to ~1.1 kW $cm^{-2}$. Uniform epi-illumination of a field of view of ~10-30 μm in diameter was achieved by under-illuminating the back aperture of a plan-apochromatic TIRF 100x, 1.49 N.A. oil immersion objective (Nikon, Japan) mounted on a piezo objective Z-axis scanner (P-725 PIFOC, Physik Instrumente, Karlsruhe, Germany). A quarter-wave plate (375/550 nm, B. Halle Nachfolger GmbH, Berlin, Germany) was inserted into the illumination path to ensure nearly circular polarization of the laser beam. The fluorescence emitted by the sample was collected by the same objective lens and separated from the exciting laser light by a dichroic filter (Z580dcxr, AHF Analysentechnik, Tübingen, Germany). The 100x objective position was controlled along the Z-axis through a P-721.11 PIFOC nano-positioner and an E-662 objective controller from Physik Instrumente, Karlsruhe, Germany. Residual excitation laser light was removed by a notch filter (NF01-532U-25, AHF Analysentechnik, Tübingen, Germany) and the detection range was limited to the label emission spectrum by a single-band bandpass filter (for AlexaFluor[555], we used a FF01-586/20 filter, Semrock, Rochester, NY). For Texas Red, the Nikon filter cube AT-TRITC LP filter cube was employed. The emission beam was magnified by 1.5x and imaged onto the back-illuminated electron multiplying CCD camera (iXon-Plus DU-860, Andor Technology, Belfast, Northern Ireland) driven at a gain of ~250, with a final pixel size of 106 nm, 3.6 photoelectrons per A/D count.

**Cell-surface fluorescence staining of nAChRs**
CHO-K1/A5 cells grown on 18 mm diameter No. 1.5 glass coverslips (WRL) in Ham's F12 medium at 37°C were washed thrice with M1 medium, incubated in M1 medium for 45 min-1 h at 4°C with Alexa[555]-BTX at a final concentration of 1 µM and finally washed thrice with cold M1. Coverslips with the adhered cells were subsequently mounted in open holder chambers built at the Max-Planck-Institute for Biophysical Chemistry in Göttingen, Germany.

**Single-molecule superresolution imaging of live cells**
STORM imaging buffer[6, 7] consisting of 1% v/v glucose (500 mg/mL stock solution), 1% v/v glucose oxidase (5000 U/mL stock), 1% v/v catalase (40,000 U/mL stock) and 0.5% v/v 2-mercaptoethanol was filtered, degassed and UV irradiated under a transgel illuminator for 20 min prior to use to diminish background fluorescence. Coverslips with the cells stained with fluorescent BTX as described above were mounted in the custom-designed chambers for imaging at room temperature. The M1 physiological saline was replaced by the STORM imaging buffer just immediately before imaging proper. Cell viability is reported to be maintained for at least 20 min in this type of buffer[8]. Streams of single frames were acquired within the time window of <8 min as described below. All other imaging steps were carried out as described for fixed specimens. Cells were inspected after image acquisition to ensure preservation of cell morphology.

**Single-molecule superresolution imaging of fixed specimens**
CHO-K1/A5 cells were stained with mAb35 and subsequently with Alexa[555]- or Texas Red-labelled goat anti-mouse secondary antibody for 1 additional h at 4°C, and fixed with 2% paraformaldehyde containing 2% sucrose for 20 min at room temperature, washed thrice with M1 supplemented with 0.1M glycine containing 10% bovine foetal serum, covered with 20 µL of a 1% aqueous solution of 25,000 MW PVA in Millipore-filtered distilled water, spun in a table top centrifuge rotor shaft to form a thin layer of PVA [9], and finally mounted onto the holder chambers. Alternatively, a drop of Prolong Gold was added to the coverslip and the latter was allowed to settle overnight onto a glass slide. Cells were initially inspected with LED excitation under low intensity illumination conditions and appropriate areas were selected for STORM imaging. Illumination was then switched to the high-intensity laser source, and the CCD camera was set to acquire a stream of images at maximum frame rate. Between 7,000 and 10,000 frames were acquired at rates of ~10 ms/frame from the ventral, coverslip-adhered surface of 8-15 cells for each experimental condition with an Andor iXon Plus DU-860 EM-CCD camera with a 1.0 or a 1.5x projection lens, yielding a pixel size of 160 nm or 106 nm in the image plane, respectively. Streaming movies were acquired using the software SlideBook (Intelligent Imaging Innovations, Boulder, CO) and exported as 16-bit TIF or Matlab files for subsequent off-line analysis.

**Superresolution data analysis**

**i) Determination of sub-diffraction molecular coordinates**

The off-line localization of the x,y coordinates of the nAChR spots was carried out using the image analysis package ThunderSTORM (https://code.google.com/p/thunder-storm/) [10] run as a plugin in ImageJ (https://imagej.nih.gov/ij/). ThunderSTORM is particularly suitable for separating multiple overlapping PSFs (typical emitter density was 3.78 ± 0.01; see Suppl. Fig. 1c). To account for the discrete nature of pixels in digital cameras, an integrated form of a symmetric 2D Gaussian function was fitted to the spots using Levenberg–Marquardt least-squares minimization routines. Localizations that were too close together to be independent were discarded. The ThunderSTORM multi-emitter fitting analysis was enabled, and the limiting intensity range was set at 500-2,000 photons. Other ThunderSTORM filters were enabled to remove uncertainty-based duplicates (e.g. multiple emitters, duplicate localizations, as described in ref. [11]). Lateral drift was estimated experimentally using fiducial 100 nm coverslip-adhered fluorescent beads and corrected via the appropriate filter in ThunderSTORM. Localization precision was calculated automatically via ThunderSTORM using a modified version of the formula in ref. [12] which considers the EM gain of the EM-CCD camera (Quan et al. 2010), through the following expression:

$$\langle (\Delta x)^2 \rangle = \frac{2\sigma^2 + a^2/12}{N} + \frac{8\pi\sigma^4 b^2}{a^2 N^2} \qquad (1)$$

Where $\sigma$ is the standard deviation of the fitted point spread function (PSF), $a$ is the pixel size in nm, $N$ is the intensity expressed in number of photons and $b$ is the background signal level in photons. The average localization precision was 37.58 ± 0.02 nm (see Suppl. Fig. 1a).

**ii) Single-particle tracking (SPT)**

The fluorescent particles were detected by a generalized likelihood ratio test algorithm specifically designed to detect point spread function-shaped (i.e. Gaussian-like) spots using ThunderSTORM [10] and exported in a format suitable for tracking analysis using an ad-hoc Matlab routine written in our laboratory. Detected (validated) particles were further analyzed for their trajectories with the software package Localizer (https://bitbucket.org/pdedecker/localizer)[13] implemented in Igor Pro (Wavemetrics Inc. https://www.wavemetrics.com). Two critical parameters were set in Localizer: the maximal number of frames (3 = 30 ms) that a given molecule was allowed to blink ("Max blinking"), and the maximum distance ("Max jump distance") at which two points could lie and be attributed to the same trajectory (3 pixels). The optimal "Max blinking" ($t_{off}$) was determined following the method of Annibale and coworkers [14], which implies that the number of photoblinking fluorescent molecules $N$ in the sample can be estimated from the number of counts (localizations) at different dark times $t_d$, $N(t_d)$, by fitting to the semi-empirical equation:

$$N(t_d) = N\left(1 + n_{blink} e^{\left(\frac{1-t_d}{t_{off}}\right)}\right) \quad (2)$$

in the regime of low dark time $t_d$ values [14]. The maximum distance for the merging filter in our experiments was determined from the camera pixel size and the distance-filter criterion described by others[15]. Briefly, localized molecules that reappeared in consecutive frames were considered as corresponding to the same molecule if the frame-to-frame displacement (tracking radius) was within 106 nm, thus allowing the monitoring of molecules with diffusion coefficients of up to 1.33 µm² s⁻¹, i.e. conservatively higher than the upper bound for nAChR nanocluster diffusion estimated from TIRF-SPT experiments in our laboratory[3]. Following the above argument, the "Max jump distance" was deduced by considering the maximum distance that a nAChR is allowed to "jump" within the merge time window established above, i.e. the maximum distance that, on average, a single-molecule can travel in 3 frames.

### iii) Mean-square displacement (MSD) analysis
Typically, a diffusion process is characterized by the mean-square displacement (MSD), which for a 2-dimensional space like a membrane bilayer can be written as:

$$\text{MSD} = <\Delta r^2(t)> = \int_{-\infty}^{\infty} r^2(t) P(r,t)\, d^2r = 4Dt \quad (3)$$

Where *D* is the diffusion constant. This assumes a purely viscous and homogeneous fluid, such that $P(r,t)$ is the probability distribution function (PDF, also termed propagator) of the diffusion process, i.e., the probability of finding the particle at a (radial) distance *r* away from the origin at time *t* after release of the particle at *r* = 0 at time *t* = 0.

Complex media may lead to sublinearity of the MSD as a function of time:

$$\text{MSD} = <\Delta r^2(t)> = K_\beta t^\beta \quad (4)$$

In equation 3, anomalous diffusion is taken into account by introduction of the exponent $\beta$ [16, 17], where $(MSD) \sim t^\beta$. Whereas for $\beta = 1$ simple thermally-driven ("random walk") Brownian diffusion results, two forms of anomalous diffusion result from other values of $\beta$: subdiffusion for $0 < \beta < 1$ (e.g. in molecular crowding), and superdiffusion for $\beta > 1$. Thus, in the case of anomalous diffusion Eq. (4) above can be written in short-form as: $4Dt^\beta$.

The time-averaged mean-square displacement (tMSD) was calculated for each trajectory *j* in the form:

$$\text{tMSD}\left(t_{lag} = m\Delta t\right) = <\Delta r^2\left(t_{lag} = m\Delta t\right)>_T = \frac{1}{M}\sum_{i=1}^{T/\Delta t} R_j^2(t_i + m\Delta t) \quad (5)$$

$$R_j^2(t_i + m\Delta t) = [x_j(t_i + m\Delta t) - x_j(t_i)]^2 + [y_j(t_i + m\Delta t) - y_j(t_i)]^2 \quad (6)$$

where $(x_j, y_j)$ is the position sampled at *M* discrete times $t_i = i\Delta t$ with displacements different from NaN, $\Delta t$ is the acquisition time (in our case, $\Delta t = 10$ ms), T is the total averaging time and *i* is the frame number.

The ensemble-averaged mean-square displacement (eMSD) was calculated over a time interval $m\Delta t$,

$$\text{eMSD}\left(t_{lag} = m\Delta t\right) = <\Delta r^2\left(t_{lag} = m\Delta t\right)>_{ens} = \frac{1}{N}\sum_{j=1}^{N} R_j^2(t_i + m\Delta t) \quad (7)$$

where *N* is the total number of available single-molecule trajectories (not NaN) at time $t_i$, where $t_i$ is the starting time relative to the first point in the trajectory. Here we follow the nomenclature employed by Krapf and coworkers and the calculation of the tMSD and eMSD was done following their procedures[18]. Following the criteria of ref.[19] the power (anomalous) exponent $\beta$ was obtained by linear fitting the initial 50 points of the log-log transformed tMSD and the eMSD, respectively. The generalized diffusion coefficient, $K_\beta$, was obtained from the linear fit to the first 50 time points of the individual tMSDs in log-log scale, evaluated at $t = 1$ (see Eq. 4), also following the criteria of ref.[19].

**iv) Exclusion of immobile particles in the analysis of nAChR trajectories**

Stationary (immobile) molecules were excluded from the analysis of single-molecule trajectories following a series of recently introduced criteria[19]. The procedure sets a threshold value on the ratio of the radius of gyration $R_g$ and the mean step size $|\Delta r|$ of the particles´ displacement. In the case of ideal immobile particles this ratio is constant, whereas for mobile particles the ratio increases. The normalized ratio:

$$(\sqrt{\pi/2} \ (R_g/\langle|\Delta r|\rangle)) \qquad (8)$$

was obtained from experiments with paraformaldehyde-fixed cells, and the ratio was subsequently employed to obtain the threshold value applied to live cells experiments. Golan and Sherman (2017) discuss the advantages of this method over the use of the diffusion coefficient or $R_g$ alone for excluding immobile particles; the two latter procedures would falsely classify immobile particles as mobile. Threshold values with >95% confidence were obtained by pooling data from different cells in independent sets of experiments.

**v) Classification of mobile particles according to their diffusivity and ergodicity analysis**

To categorize mobile trajectories into subpopulations according to their diffusivity, we first plotted the tMSDs in a log-log scale. Linear fitting of the curves rendered the power (anomalous) exponent β and the goodness of the fit. Those having a goodness of fit better than 0.9 were selected. The anomalous exponent β was then used to classify trajectories into 5 arbitrary subpopulations: subdiffusive I (β < 0.5), subdiffusive II (0.5 ≤ β< 0.7), subdiffusive III (0.7 ≤ β< 0.9), Brownian (0.9 ≤ β < 1.1), and superdiffusive (β ≥ 1.1).

In statistical mechanics, the ergodic theory predicts that for large systems of interacting particles at equilibrium, the time average along a single trajectory equals the ensemble average (i.e. for sufficiently long measurement times, the time average provides the same information as the ensemble average). Thus, the observed diffusion coefficient obtained from an individual trajectory is identical to the diffusion constant found from an ensemble of particles under identical physical conditions. This equivalence between the ensemble diffusion and the behaviour of an individual representative particle indicates that the process is ergodic (see recent review and references therein in ref.[20]). When inequality holds at long measuring times it is assumed that the process violates the Boltzmann/Khinchin ergodic hypothesis[21,22]. To determine whether the nAChR motion at the cell surface was ergodic or not, the time- and ensemble-averaged mean-square displacements (tMSD and eMSD, respectively) were calculated for each experimental condition and the anomalous exponent β was obtained as described above.

### vi) Escape time distributions

The escape (waiting) time is the interval during which a trajectory remains within a given radius $R_{TH}$. The duration of the events in which the molecules' trajectories remained within circular areas of increasing radii $R_{TH}$ was calculated as in ref.[18]. Briefly, for each trajectory $j$:

a) We generated the displacement squares

$$R_{t_{lag}=\Delta t,i,j}^2 \qquad (9)$$

for each time $t_i = i\Delta t$, where $\Delta t$ corresponds to the inter-frame time (in our case, $\Delta t = 10$ ms).

b) Next, we looked for the first displacement for which

$$R_{t_{lag}=\Delta t,i,j}^2 < R_{TH}^2 \qquad (10)$$

We called this time $t_k = k\Delta t$, i.e. the time corresponding to the frame $k$ at lag time $t_{lag} = \Delta t$.

c) Next, we observed the displacement squares at times $t_k$ for increasing lag times $t_{lag} = 2\Delta t, 3\Delta t, \ldots$ until either the relation $R_{t_{lag},k,j}^2 > R_{TH}^2$ was satisfied or the trajectory ended. This is true for a certain escape time

$$t_{lag} = t_F = F\Delta t \qquad (11)$$

d) When the escape time was obtained, we continued from $t_k + t_F + \Delta t = t_{k+F+1}$ looking for the next displacement that satisfied the relationship:

$$R_{\Delta t,i,j}^2 < R_{TH}^2 \qquad (12)$$

for $(i \geq k + F + 1)$

e) If the step was found, steps (b-c) were repeated until the end of the trajectory.

Finally, we constructed the cumulative distribution function for all the resulting escape times.

### vii) Turning angle (directional change) analysis

The turning angle distribution as a function of the time lags has been employed to search for correlations in the particles' displacements. We applied this analysis to follow the directional changes in individual nAChR trajectories using the relative angles distended by the molecules along their walk, a parameter which can help distinguish among different types of anomalous subdiffusive mechanisms, among them, the so-called fractional Brownian motion (fBM) and obstructed diffusion models[23].

### viii) Single-molecule trajectory recurrence analysis

The presence of confinement periods within individual trajectories was identified using recently developed algorithms[24], based on the evaluation of the total number of visits (recurrence) performed by the moving particle to a given site. Confinement within a nanoscale domain is associated with multiple visits to the same site, multiple times, in a short period, whereas unconfined motion is related to exploration of wider regions, less compact walks, and less visits to the sites previously walked. The analyses characterize subpopulations of trajectories based on the times spent in each state and the areas distended in confinement.

### ix) Identification of dynamic nAChR nanoclusters

We applied the centroid-linkage hierarchical clustering and the density-based spatial clustering with noise (DBSCAN) analyses embedded in the open source qSR software developed by Cissé and coworkers[25] (www.github.com/cisselab/qSR) to identify in real time the formation and breakage of clusters of nAChR molecules. The length scale was set at 100 nm, and the minimal number of particles was 20 in the DBSCAN analysis. Dark-time tolerance was set at 1 s. All localizations were included in this analysis, and only for this purpose.

**Statistical analyses**

These were done using one-way analysis of variance employing ANOVA or Kruskal-Wallis, where appropriate. Multiple comparisons tests were performed using Tukey's or Dunn's approaches implemented in the Prism GraphPad software. Mean ± 95% confidence intervals are shown unless otherwise stated. The one-sample Kolmogorov-Smirnov test was applied to assess whether the data were normally distributed or not. To compare two distributions, we used the Kolmogorov-Smirnov (KS) test for two samples.

# Results

**Localization precision and merge time**

The localization precision and merge time are two essential parameters in STORM and in single-molecule localization microscopy in general. Suppl. Figure 1 shows the histograms for determining the localization precision as applied to BTX-labelled nAChR molecules.

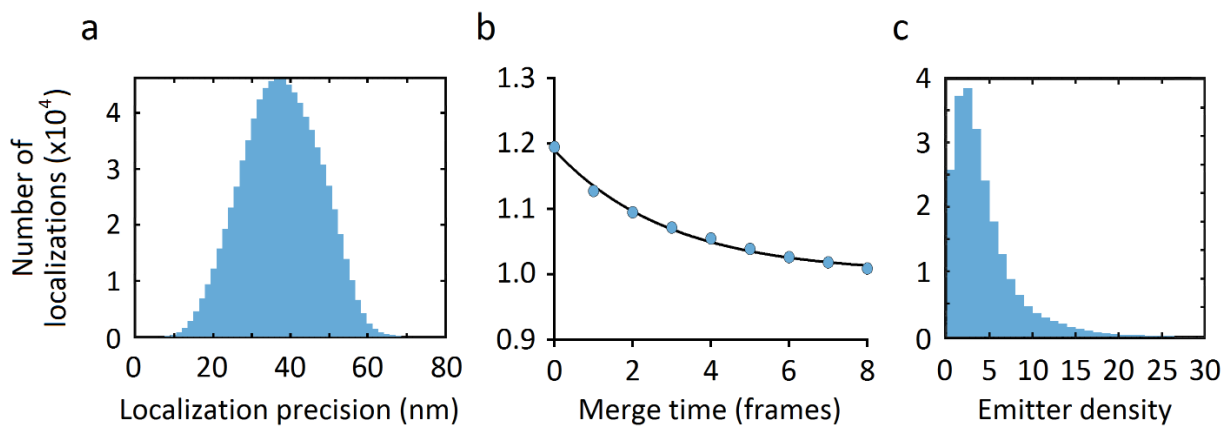

**Supplementary Figure 1. Localization precision and merge time**. a) Histogram of the localization precision using the method of Quan and coworkers[26] (see eq. (1) ) as applied to two CHO-K1/A5 cells labelled with BTX. The average localization precision was 37.58 ± 0.02 nm. b) Determination of the optimal merge time following the method of Annibale and coworkers [14]. The plot shows the total number of localizations against the merge time in a representative image of BTX-labelled nAChRs and the corresponding fit to the semi-empirical Eq. (2). In the example shown, the optimal merge time was found to be three frames (30 ms). c) Histogram of the number of localized molecules per frame, i.e. the emitter density. The typical emitter density was 3.78 ± 0.01.

**Classification of nAChR tracks into mobile and immobile trajectories.**

Suppl. Figure 2 shows the ratio of the radius of gyration $R_g$ and the mean step size (a), examples of mobile and immobile particles (b), and the experimentally determined percentage of immobile particles upon application of the Golan and Sherman's criteria[19]. The combination of the radius of gyration and the mean displacements of nAChR validated localizations in paraformaldehyde-fixed cells (Suppl. Figure 2a) led us to set a threshold value of 2.1 which was used as the cutoff for exclusion of immobile particles. This enabled us to calculate the relative proportion of mobile/immobile particles in live speciments (Suppl. Figure 2), and exclude immobile molecules from further analysis. The CHO-K1/A5 mammalian clonal cell line lacks receptor-immobilizing molecules like rapsyn or clustering non-receptor scaffolding proteins like agrin and MusK. Immobilization must therefore respond to other scaffolding molecules or the self-aggregation of the nAChR protein in higher oligomeric species[27], or a combination thereof.

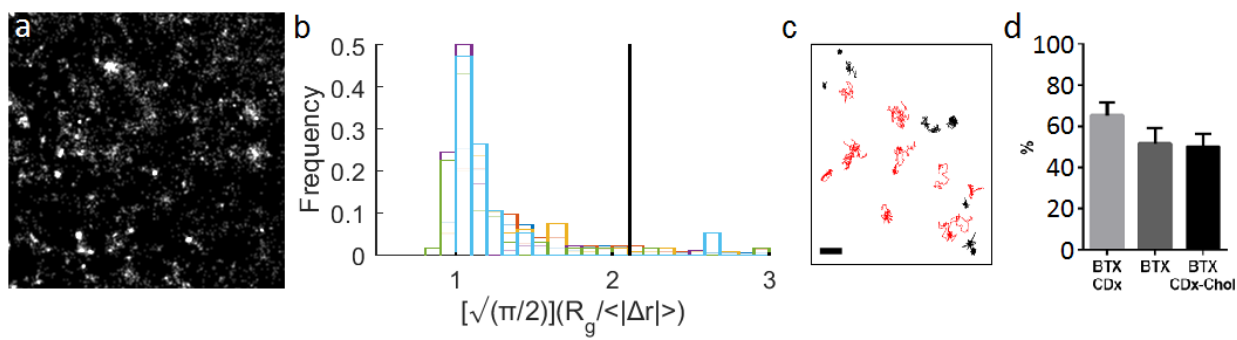

**Supplementary Figure 2. Criteria for classification of nAChR trajectories into mobile and immobile.** a) A 5 x 5 μm Gaussian-smoothened projection image of a fixed CHO-K1/A5 cell stained with mAb35 followed by Texas Red-labelled goat secondary antibody, displaying the single-molecule localizations. b) Distribution of the ratio of the radius of gyration and the mean step size for two independent samples of nAChRs in paraformaldehyde-fixed cells (n=6). Threshold values of 1.5-2.1 (95% confidence) were obtained[19] and the value of 2.1 was chosen (black line). c) Mobile (red) and immobile (black) nAChR trajectories under control conditions resulting from application of the criteria of ref.[19] for separating particle walks into these two categories. Scale bar: 500 nm. d) Comparison of the percentages of immobile nAChR trajectories upon application of the threshold resulting from Suppl. Figure 2a. Bars indicate mean ± S.E.M.

**Modification of cholesterol levels of CHO-K1/A5 cells**

Methyl-β-cyclodextrin (CDx) is a widely used tool to acutely modify the cholesterol levels at the plasma membrane. The kinetics and thermodynamics of cholesterol efflux from /incorporation into the membrane are relatively well understood: once CDx establishes contact with the plasmalemma, cholesterol migrates from the membrane to the CDx core [28] [29]. We have used CDx alone, or in complex with cholesterol, to deplete (CDx) or replenish (CDx-Chol), respectively, the cell-surface sterol content and assess cholesterol effects on receptor transport mechanisms[30, 31], nAChR diffusion[3, 32, 33] and cell-surface organization[34]. Here we applied CDx or CDx-Chol and analysed the effects on the nAChR at the ensemble level or at the level of trajectories divided into subpopulations according to diffusivity. Suppl. Table 1 below lists the percentage of trajectories found in each subpopulation under control and cholesterol modifying conditions.

**Supplementary Table 1. Percentage of trajectories in subpopulations separated according to their power (anomalous) exponent β* under control and cholesterol-modifying conditions****

| Treatment | Subdiffusive I | Subdiffusive II | Subdiffusive III | Brownian | Superdiffusive |
|---|---|---|---|---|---|
| CDx | 10.75 ± 7.64 | 27.7 ± 8.71 | 31.04 ± 10.15 | 12.80 ± 9.74 | 3.24 ± 3.51 |
| Control | 8.63 ± 5.90 | 31.91 ± 11.65 | 30.08 ± 15.99 | 22.39 ± 13.76 | 3.08 ± 4.03 |
| CDx-Chol | 5.67 ± 8.41 | 25.79 ± 18.66 | 33.39 ± 9.84 | 20.34 ± 10.35 | 10.07 ± 8.27 |

*Groups are ordered from less diffusive (left) to more diffusive (right). Diffusivity-based classification of trajectories (see main text) rendered the following groups: subdiffusive I ($β < 0.5$), subdiffusive II ($0.5 ≤ β < 0.7$), subdiffusive III ($0.7 ≤ β < 0.9$), Brownian motion ($0.9 ≤ β < 1.1$), and superdiffusive ($β ≥ 1.1$). Values are the mean ± S.D. Statistically significant differences between BTX control and BTX CDx-Chol superdiffusive ($p=0.042$). **Trajectories not satisfying a goodness of fit of 0.9 are not included in this Table. They represented a variable proportion of the total, amounting to ~4-14%.

**Time-averaged mean-square displacement (tMSD) of nAChR trajectories**

The mean-square displacements (MSDs) of the mobile trajectories were analysed in the total unsorted population of nAChRs (Figure 2 in main text). Their ergodic and non-ergodic behaviour can be appreciated from inspection of the values of the anomalous exponent β (Table 1 in main text). The plots in Supplementary Figure 3 below illustrate the MSD curves for the subpopulations separated according to diffusivity.

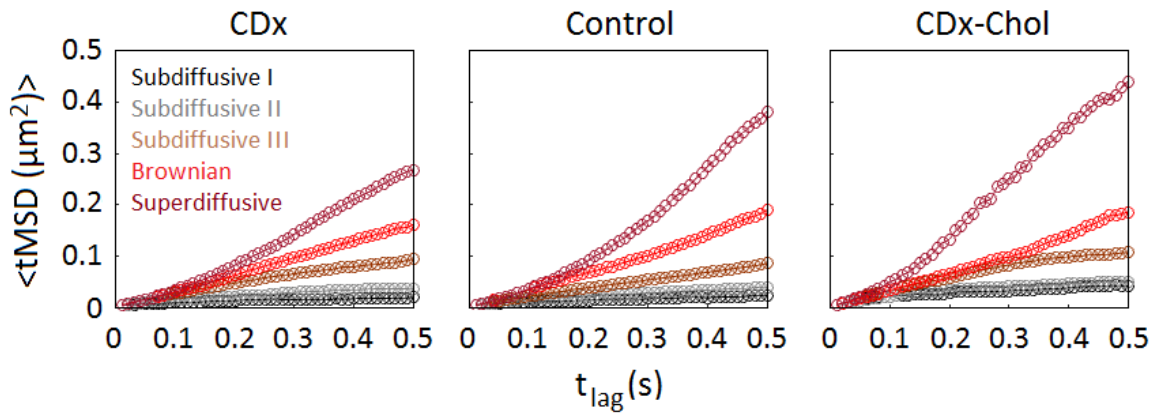

**Supplementary Figure 3. Average time-averaged mean-square displacement (tMSD) of nAChR trajectories separated into subpopulations according to their power (anomalous) exponent β.** Log-log plots of the average tMSD having a goodness of fit better than 0.9 (see Material and

Methods). Each colour corresponds to a different diffusivity-based subpopulation ordered according to its power exponent.

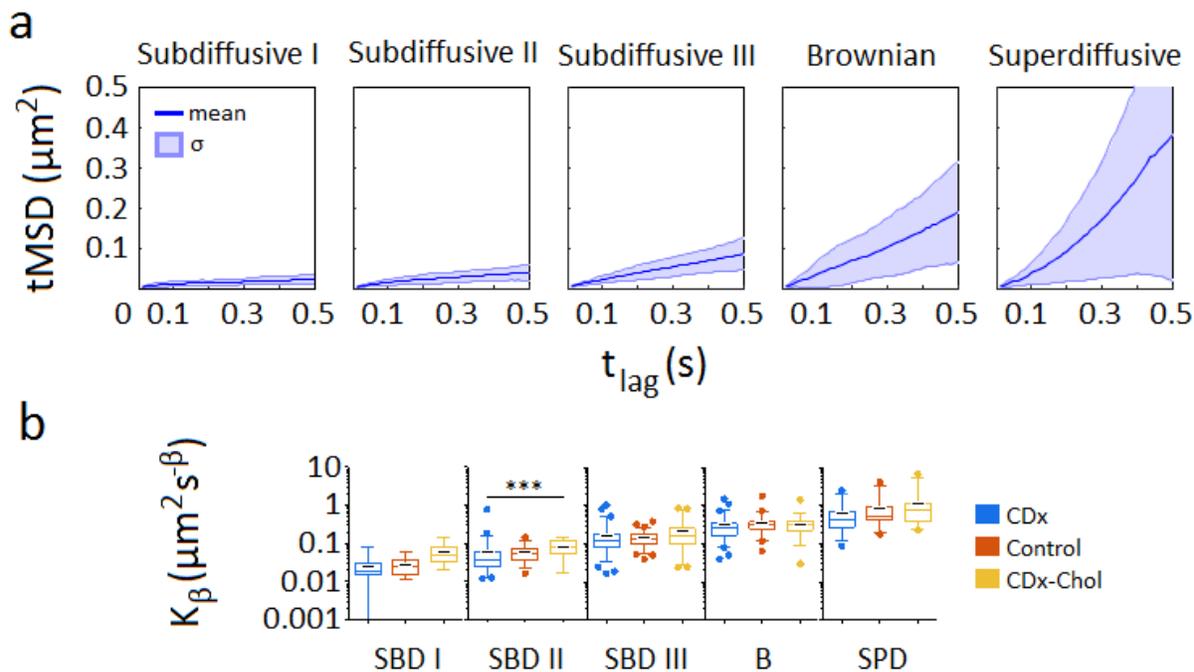

**Supplementary Figure 4.** a) Average tMSDs of nAChR trajectories separated according to diffusivity. b) Generalized diffusion coefficients, $K_\beta$, of the subpopulations of trajectories classified according to their anomalous exponent β (plots are ordered according to increasing β ranges as shown above in (a)). Whiskers in box plots correspond to 95% confidence intervals. The limits indicate 75% confidence intervals; the black negative symbols indicate the mean and the horizontal lines the median in each case. The dots outside the confidence intervals are outliers. Statistics: (***), p<0.001.

**Turning angle analysis**.
Turning angle analysis was recently employed to study the correlation of experimentally determined single-molecule steps in voltage-gated potassium channels Kv1.4 and Kv2.1 [35] in comparison to numerical simulations of fractional Brownian motion (fBM) and obstructed diffusion (OD) models. A schematic diagram of this type of analysis is shown in Suppl. Figure 5. Experimental results corresponding to the cholesterol-modified nAChR subpopulations separated according to diffusivity are shown in Suppl. Figure 6.

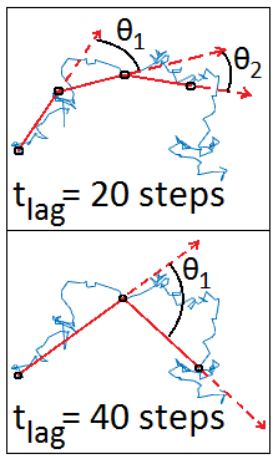

**Supplementary Figure 5.** Schematic diagram depicting the procedure in turning angle analysis [23]. At the starting position the observer follows the trajectory (blue trace) with a given number of steps (20 or 40 in the example shown) defining the time-lag ($t_{lag}$), and the angles ($\theta_1$, $\theta_2$….) distended by successive steps.

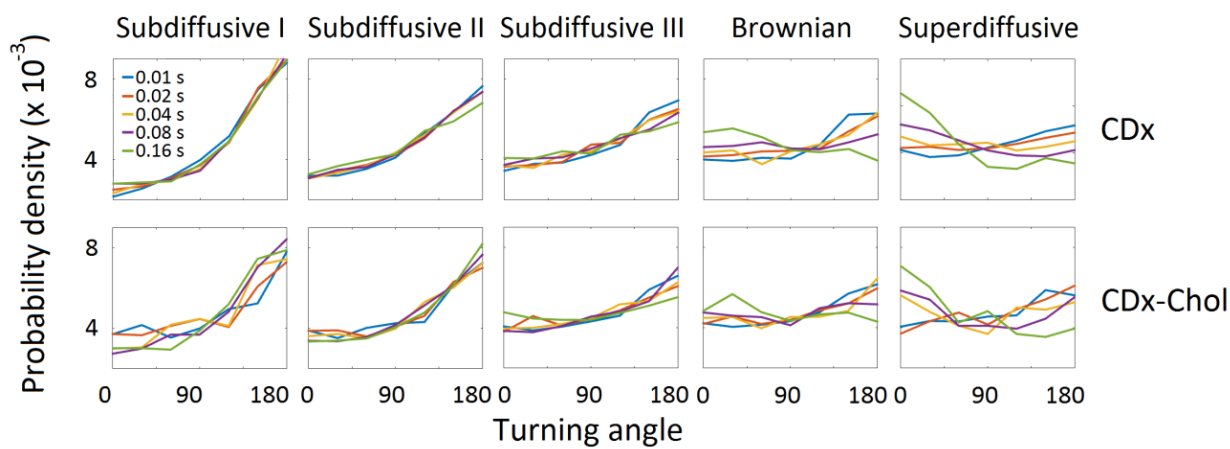

**Supplementary Figure 6. Turning angle probability densities for the diffusivity-classified trajectories under cholesterol depletion (CDx) and cholesterol-enrichment (CDx-Chol) conditions.** Probability densities for the colour-coded $t_{lag}$ of increasing durations (10 to 160 ms). The probability density is normalized such that the integral of the curve is equal to unity.

**Transient confinement nanodomains within individual trajectories and differences between subdiffusive states**

The results of the recurrence analysis[36] as applied to the individual trajectories in the diffusivity-based subpopulations under control conditions are shown in the main text (Figure 5). This analysis disclosed microscopic heterogeneity in all trajectories and differences in diffusion characteristics among subpopulations. Suppl. Figure 7a shows the differences in the areas transiently occupied by the trajectories in their confined portion. The areas covered by the confined section of the trajectories show statistically significant differences between all the subdiffusive subpopulations (which are similar to each other) and the Brownian trajectories ($p < 0.0001$). The cumulative distribution times (Suppl. Figure 7b) show differences between different subpopulations, which are attenuated for the faster diffusional regimes. In the subdiffusive I subpopulation, a marked shortening of the characteristic decay time is observed upon cholesterol modification (Supplementary Table 2). This is in marked contrast with the Brownian subpopulation, in which case cholesterol modification appears to lengthen the residence time in the confined state (Supplementary Table 2).

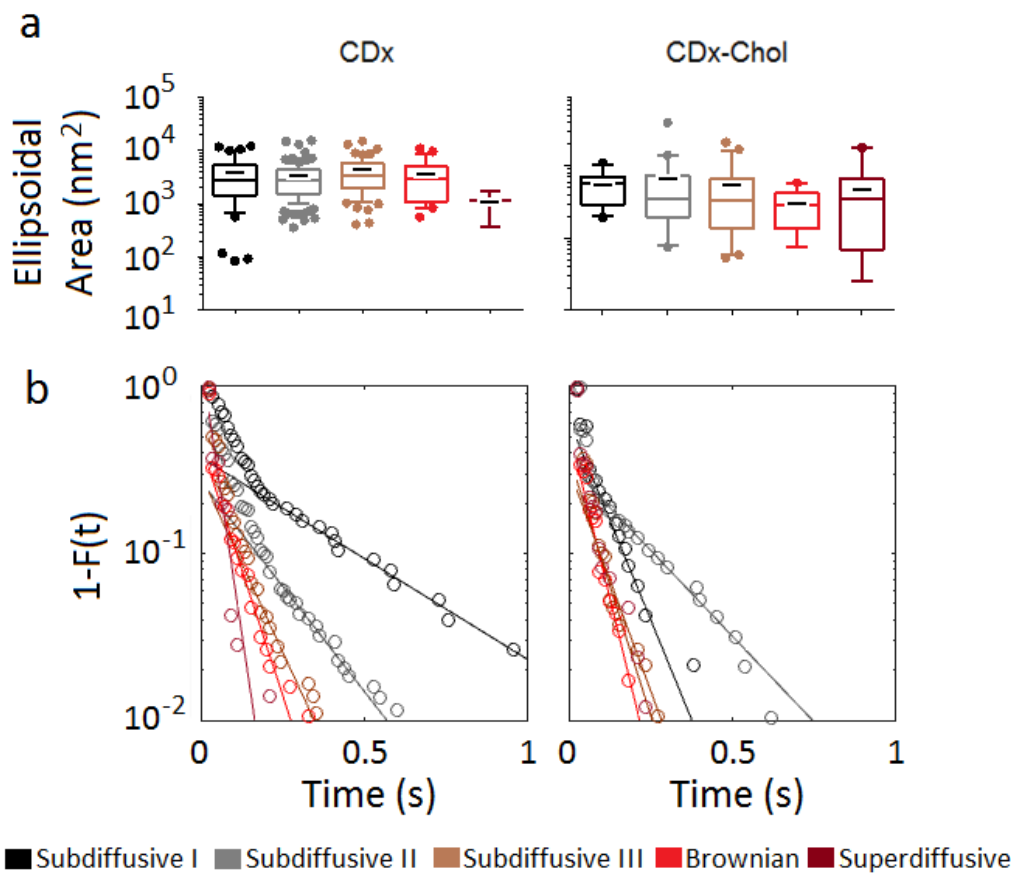

**Supplementary Figure 7. Confinement areas and complementary cumulative distribution function of the residence times in the confined state under cholesterol modification conditions for each diffusivity-based group.** a) Mean ellipsoidal area of the trajectories´ portion in the confined state. b) The sojourn times have an exponential distribution tail with a characteristic decay time, which turns out to be inversely proportional with the subpopulation diffusivity (see Suppl. Table 2 below). Whiskers in box plots correspond to 95% confidence intervals. The limits indicate 75% confidence intervals; the black negative symbols indicate the mean and the horizontal lines the median in each case. The dots outside the confidence intervals are outliers.

**Supplementary Table 2. Decay times of the confined portions corresponding to the individual trajectories in subpopulations separated according to their power (anomalous) exponent β and in the entire population*.**

| Subpopulation | Subdiffusive I | Subdiffusive II | Subdiffusive III | Brownian | Superdiffusive |
|---|---|---|---|---|---|
| CDx | | | | | |
| Tau (s) | 0.361 ± 0.033 | 0.174 ± 0.009 | 0.103 ± 0.009 | 0.073 ± 0.007 | 0.033 ± 0.013 |
| Control | | | | | |
| Tau (s) | 0.310 ± 0.022 | 0.129 ± 0.012 | 0.116 ± 0.008 | 0.041 ± 0.003 | 0.045 ± 0.014 |
| CDx-Chol | | | | | |
| Tau (s) | 0.097 ± 0.044 | 0.213 ± 0.012 | 0.060 ± 0.012 | 0.044 ± 0.008 | 0.055 ± 0.008 |

| Total Population | CDx | Control | CDx-Chol |
|---|---|---|---|
| Tau (s) | 0.257 ± 0.007 | 0.171 ± 0.009 | 0.135 ± 0.005 |

The question arose as to whether the different diffusional regimes are given by the concentration of obstacles -as suggested by the progressive decrese in the slope of the turning angles (Figure 4b and Suppl. Figure 6) or, in contrast, whether they result from the averaging of the lifetimes of confined and free regions of the individul trajectories. Suppl. Figure 8 below shows that the slope of the turning angle in the confined portion of the individual trajectories is essentially the same, confirming the idea that the concentration of obstacles is the same for all the confined regions independently of the ensemble diffusional regime; it is the time spent in one state or the other which determines the diffusional modality, as supported by the clear correlation between the diffusivity-based subpopulations with the percentage of single-molecule confinement and the subsequent disappearance of subdiffusive motifs in the free portions of the trajectories, as shown in Supplementary Figure 9 below.

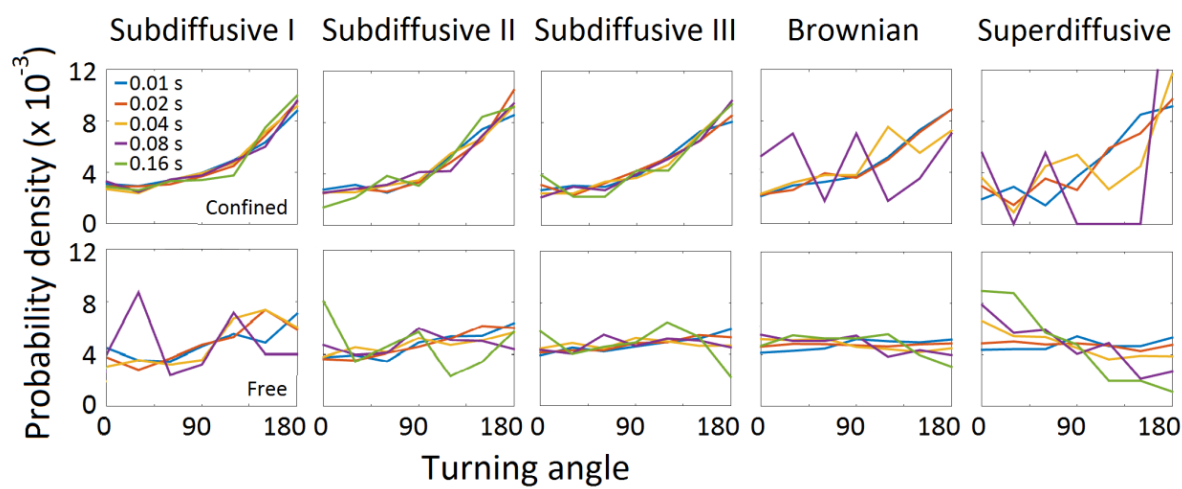

**Supplementary Figure 8. Turning angle probability densities of the transiently confined and unconfined portions of the individual trajectories under control conditions.** Probability densities for the colour-coded $t_{lag}$ of increasing durations (10 to 160 ms). The probability density is normalized such that the integral of the curve is equal to unity.

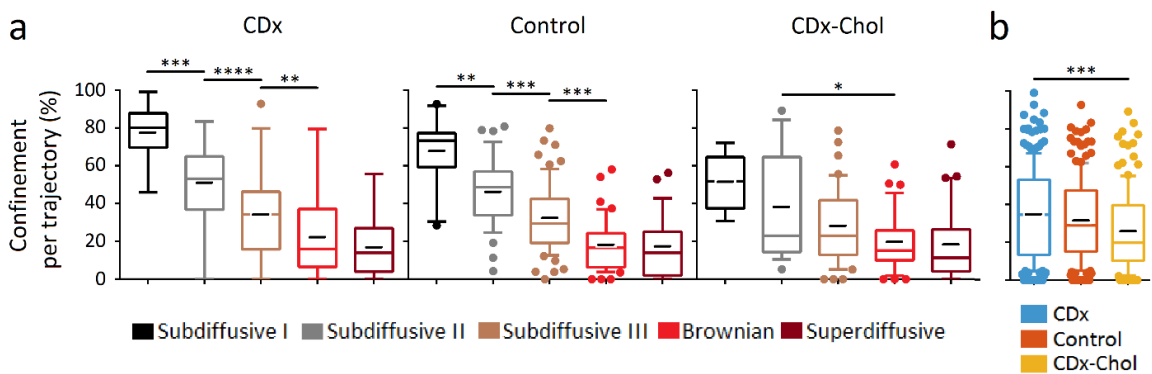

**Supplementary Figure 9. Percentage of confinement events in diffusivity-classified single-molecules and anomalous exponent β distributions in free and confined portions of the total population of trajectories.** a) Confinement per single-molecule trajectory in diffusivity-based subpopulations and b) for the entire population. Whiskers in box plots correspond to 95% confidence intervals. The limits indicate 75% confidence intervals; the black negative symbols indicate the mean and the horizontal lines the median in each case. The dots outside the confidence intervals are outliers. p-values: (*), p<0.05; (**), p<0.01; (***), p<0.001 and (****), p<0.0001.

**References**
1. Roccamo, A.M. *et al.* Cells defective in sphingolipids biosynthesis express low amounts of muscle nicotinic acetylcholine receptor. *The European journal of neuroscience* **11**, 1615-1623 (1999).
2. Borroni, V. *et al.* Cholesterol depletion activates rapid internalization of submicron-sized acetylcholine receptor domains at the cell membrane. *Mol.Membr.Biol.* **24**, 1-15 (2007).
3. Almarza, G., Sanchez, F. & Barrantes, F.J. Transient cholesterol effects on nicotinic acetylcholine receptor cell-surface mobility. *PloS one* **9**, e100346 (2014).
4. Rust, M.J., Bates, M. & Zhuang, X. Sub-diffraction-limit imaging by stochastic optical reconstruction microscopy (STORM). *Nat.Methods* **3**, 793-795 (2006).
5. Folling, J. *et al.* Fluorescence nanoscopy by ground-state depletion and single-molecule return. *Nat.Methods* **5**, 943-945 (2008).
6. Harada, Y., Sakurada, K., Aoki, T., Thomas, D.D. & Yanagida, T. Mechanochemical coupling in actomyosin energy transduction studied by in vitro movement assay. *J Mol Biol* **216**, 49-68 (1990).
7. Adachi, K. *et al.* Stepping rotation of F1-ATPase visualized through angle-resolved single-fluorophore imaging. *Proc Natl Acad Sci U S A* **97**, 7243-7247 (2000).
8. Jones, S.A., Shim, S.H., He, J. & Zhuang, X. Fast, three-dimensional super-resolution imaging of live cells. *Nat. Methods* **8**, 499-508 (2011).
9. Barrantes, F.J. Single-molecule localization superresolution microscopy of synaptic proteins., in *Springer Protocols*, Edn. 2016. (ed. A.K. Shukla) 1-42 (Springer Science+Business Media, 2016).
10. Ovesný, M., Krizek, P., Borkovec, J., Svindrych, Z. & Hagen, G.M. ThunderSTORM: a comprehensive ImageJ plug-in for PALM and STORM data analysis and super-resolution imaging. *Bioinformatics (Oxford, England)* **30**, 2389-2390 (2014).
11. Huang, F., Schwartz, S.L., Byars, J.M. & Lidke, K.A. Simultaneous multiple-emitter fitting for single molecule super-resolution imaging. *Biomedical optics express* **2**, 1377-1393 (2011).
12. Thompson, R.E., Larson, D.R. & Webb, W.W. Precise nanometer localization analysis for individual fluorescent probes. *Biophys.J* **82**, 2775-2783 (2002).
13. Dedecker, P., Duwé, S., Neely, R.K. & Zhang, J. Localizer: fast, accurate, open-source, and modular software package for superresolution microscopy. *Journal of biomedical optics* **17**, 126008-126008 (2012).
14. Annibale, P., Vanni, S., Scarselli, M., Rothlisberger, U. & Radenovic, A. Quantitative photo activated localization microscopy: unraveling the effects of photoblinking. *PloS one* **6**, e22678 (2011).
15. Lu, H.E., MacGillavry, H.D., Frost, N.A. & Blanpied, T.A. Multiple spatial and kinetic subpopulations of CaMKII in spines and dendrites as resolved by single-molecule tracking PALM. *The Journal of neuroscience : the official journal of the Society for Neuroscience* **34**, 7600-7610 (2014).
16. Tejedor, V. *et al.* Quantitative analysis of single particle trajectories: mean maximal excursion method. *Biophys.J* **98**, 1364-1372 (2010).
17. Manzo, C. *et al.* Weak ergodicity breaking of receptor motion in living cells stemming from random diffusivity. *Physical Review X* **5**, 011021 (2015).
18. Weigel, A.V., Simon, B., Tamkun, M.M. & Krapf, D. Ergodic and nonergodic processes coexist in the plasma membrane as observed by single-molecule tracking. *Proc Natl Acad Sci U S A* **108**, 6438-6443 (2011).
19. Golan, Y. & Sherman, E. Resolving mixed mechanisms of protein subdiffusion at the T cell plasma membrane. *Nature communications* **8**, 15851 (2017).
20. Metzler, R., Jeon, J.H. & Cherstvy, A.G. Non-Brownian diffusion in lipid membranes: Experiments and simulations. *Biochim Biophys Acta* **1858**, 2451-2467 (2016).
21. He, W., Song, H., Su, Y., Geng, L., Ackerson, B. J., Peng, H. B. & Tong, P. Dynamic heterogeneity and non-Gaussian statistics for acetylcholine receptors on live cell membrane. Nat. Comm. 7, 11701 (2016).

22. Cherstvy, A.G. & Metzler, R. Anomalous diffusion in time-fluctuating non-stationary diffusivity landscapes. *Physical chemistry chemical physics : PCCP* **18**, 23840-23852 (2016).
23. Burov, S. *et al.* Distribution of directional change as a signature of complex dynamics. *Proc Natl Acad Sci U S A* **110**, 19689-19694 (2013).
24. Sikora, G. *et al.* Elucidating distinct ion channel populations on the surface of hippocampal neuros via single-particle tracking recurrence analysis. *arXiv:1708.02876* (2017).
25. Andrews, J.O. *et al.* qSR: A software for quantitative analysis of single molecule and super-resolution data. *bioRxiv* (2017).
26. Quan, T. *et al.* Ultra-fast, high-precision image analysis for localization-based super resolution microscopy. *Opt Express* **18**, 11867-11876 (2010).
27. Barrantes, F.J. Oligomeric forms of the membrane-bound acetylcholine receptor disclosed upon extraction of the Mr 43,000 nonreceptor peptide. *J.Cell Biol.* **92**, 60-68 (1982).
28. Kilsdonk, E.P.C. *et al.* Cellular cholesterol efflux mediated by cyclodextrins. *J.Biol.Chem.* **270**, 17250-17256 (1995).
29. Lopez, C.A., de Vries, A.H. & Marrink, S.J. Computational microscopy of cyclodextrin mediated cholesterol extraction from lipid model membranes. *Scientific reports* **3**, 2071 (2013).
30. Pediconi, M.F., Gallegos, C.E., De Los Santos, E.B. & Barrantes, F.J. Metabolic cholesterol depletion hinders cell-surface trafficking of the nicotinic acetylcholine receptor. *Neuroscience* **128**, 239-249 (2004).
31. Borroni, V. & Barrantes, F.J. Cholesterol modulates the rate and mechanism of acetylcholine receptor internalization. *J Biol Chem* **286**, 17122-17132 (2011).
32. Baier, C.J., Gallegos, C.E., Levi, V. & Barrantes, F.J. Cholesterol modulation of nicotinic acetylcholine receptor surface mobility. *European biophysics journal : EBJ* **39**, 213-227 (2010).
33. Barrantes, F.J. Cell-surface translational dynamics of nicotinic acetylcholine receptors. *Front Synaptic Neurosci* **6**, 25 (2014).
34. Kellner, R.R., Baier, C.J., Willig, K.I., Hell, S.W. & Barrantes, F.J. Nanoscale organization of nicotinic acetylcholine receptors revealed by stimulated emission depletion microscopy. *Neuroscience* **144**, 135-143 (2007).
35. Sadegh, S., Higgins, J.L., Mannion, P.C., Tamkun, M.M. & Krapf, D. Plasma Membrane is Compartmentalized by a Self-Similar Cortical Actin Meshwork. *Physical Review X* **7**, 011031 (2017).
36. Sikora, G., Burnecki, K. & Wylomanska, A. Mean-squared-displacement statistical test for fractional Brownian motion. *Phys Rev E* **95**, 032110 (2017).